\documentclass[a4paper,12pt]{article}

\usepackage{bbm}
\usepackage[utf8]{inputenc}
\usepackage{placeins}
\usepackage{amssymb,amsmath,amsfonts}
\usepackage[colorlinks,linkcolor=purple,citecolor=teal]{hyperref}
\usepackage{dsfont}
\usepackage{color}
\usepackage{graphicx}
\usepackage{hyphenat}
\usepackage{wrapfig}
\usepackage{empheq}
\usepackage{textcomp}
\usepackage[caption=false]{subfig}
\usepackage{rotating}
\usepackage{wrapfig}
\usepackage{pdfpages}
\usepackage{verbatim}
\usepackage{setspace}
\usepackage{array}
\usepackage{caption}
\usepackage[pdf]{pstricks}
\usepackage{upgreek}
\usepackage{cmll}
\usepackage{braket}
\usepackage{cite}
\usepackage{epsfig}
\usepackage[left=2.4cm,top=3.3cm,right=2.4cm,bottom=3.3cm,bindingoffset=0cm]{geometry}
\usepackage[titletoc,page]{appendix}
\usepackage{multicol}
\usepackage{amsmath}

\usepackage{soul} 
\definecolor{mgreen}{rgb}{0, 0.3, 0.1}

\usepackage{youngtab}

\newcommand{\beq}{\begin{eqnarray}}
\newcommand{\eeq}{\end{eqnarray}}
\newcommand{\bea}{\begin{eqnarray}}
\newcommand{\eea}{\end{eqnarray}}
\newcommand{\be}{\begin{equation}}
\newcommand{\ee}{\end{equation}}

\newcommand{\diff}{\mathrm{d}}

\newcommand{\de}{\partial}

\newcommand{\tr}{\mathrm{tr}}

\newcommand{\im}{\mathrm{i}}

\newcommand{\calR}{\mathcal{R}}

\newcommand{\rmc}{\mathrm{c}}
\newcommand{\rme}{\mathrm{e}}

\newcommand{\rmA}{\mathrm{A}}
\newcommand{\rmF}{\mathrm{F}}
\newcommand{\rmL}{\mathrm{L}}

\newcommand{\rmS}{\mathrm{S}}

\newcommand{\Z}{\mathbbm{Z}}
\newcommand{\Zp}{(\Z_{N+2})_{\psi}}
\newcommand{\Zc}{(\Z_{N-2})_{\chi}}

\newcommand{\Zes}{(\Z_{N+4+p})_{\eta}}
\newcommand{\Zea}{(\Z_{N-4+p})_{\eta}}
\newcommand{\Zx}{(\Z_{p})_{\xi}}
\newcommand{\Zf}{(\Z_{2})_{F}}
\def\brc{\langle}
\def\ckt{\rangle}

\def\de{\partial}

\def\Z{\mathbb Z}
\def\1{\mathbbm{1}}

\def\S{{\cal S}}

\def\tr{\mathrm{tr}}
\def\t{\tilde}

\def\nn{\nonumber}

\def\A{{\cal A}}

\def\tr{\qopname\relax o{tr}}

\def\tU{\widetilde{U}}
\def\hU{\widehat{U}}
\def\sp{-\!\!}

\numberwithin{equation}{section}

\def\de{\partial}

\def\Z{\mathbb Z}
\def\1{\mathbbm{1}}

\def\tr{\mathrm{tr}}
\def\t{\tilde}

\def\bxi{{\bar{\xi}} }
\def\su{$ \phantom{{{{{\yng(1)}}}}}\!\!\!\!\!\!\!\!$}
\def\sbuu{$\phantom{{{{\bar{\yng(1,1)}}}}}\!\!\!\!\!\!$}
\def\sbu{$\phantom{{\bar{{{\yng(1)}}}}}\!\!\!\!\!\!\!\!$}
\def\sbbuu{$\phantom{{{\bar{\bar{\yng(1,1)}}}}}\!\!\!\!\!\!\!\!$ }
\def\sbbu{ $\phantom{{{\bar{\bar{\yng(1)}}}}}\!\!\!\!\!\!\!\!$ }

\numberwithin{equation}{section}

\begin{document}

\title{
\vskip 20pt
\bf{Probing the dynamics  of   \\    
chiral $SU(N)$ gauge theories      \\    via
generalized anomalies}
}

\vskip 60pt  

\author{  
Stefano Bolognesi$^{(1,2)}$, 
 Kenichi Konishi$^{(1,2)}$, Andrea Luzio$^{(3,2)}$    \\[13pt]
{\em \footnotesize
$^{(1)}$Department of Physics ``E. Fermi", University of Pisa}\\[-5pt]
{\em \footnotesize
Largo Pontecorvo, 3, Ed. C, 56127 Pisa, Italy}\\[2pt]
{\em \footnotesize
$^{(2)}$INFN, Sezione di Pisa,    
Largo Pontecorvo, 3, Ed. C, 56127 Pisa, Italy}\\[2pt]
{\em \footnotesize
$^{(3)}$Scuola Normale Superiore,   
Piazza dei Cavalieri, 7  56127  Pisa, Italy}\\[2pt]
\\[1pt] 
{ \footnotesize  stefano.bolognesi@unipi.it, \ \  kenichi.konishi@unipi.it,  \ \  andrea.luzio@sns.it}  
}
\date{}

\vskip 6pt
\maketitle

\begin{abstract}

We study symmetries and dynamics of  chiral $SU(N)$ gauge theories with matter Weyl fermions in a two-index symmetric  ($\psi$) or anti-symmetric tensor ($\chi$) representation,
together with  $N \pm 4 + p $ fermions in the anti-fundamental   ($\eta$)  and $p$  fermions in the fundamental  ($\xi$) representations. They are known as 
the Bars-Yankielowicz (the former) and the generalized Georgi-Glashow models (the latter).  
The conventional 't Hooft anomaly matching algorithm is known to allow a confining, chirally symmetric vacuum in all these models, with a simple set of massless baryonlike composite fermions describing the infrared physics.   

We analyzed recently one of these models ($\psi\eta$ model),   by applying the ideas of generalized symmetries and  the consequent, stronger constraints involving certain mixed anomalies, finding that  the confining, chirally symmetric,  vacuum is actually inconsistent. 

In the present paper this result  is extended to a wider class of  the Bars-Yankielowicz and the generalized Georgi-Glashow models. 
 It is shown that  for all these models with $N$ and $p$ both even,   at least,    the generalized anomaly matching requirement forbids the persistence of the full chiral symmetries in the infrared if the system confines.  The most natural and consistent possibility is that  some bifermion condensates form, breaking the color gauge symmetry dynamically, together with part of the global symmetry.

\end{abstract}

\newpage

\tableofcontents

\section{Introduction}

A few steps have been taken recently \cite{BKL1,BKL2} to go beyond the conventional 't Hooft anomaly matching analysis in understanding the infrared dynamics of 
chiral gauge theories.   The standard anomaly matching constraints and other generally accepted ideas,  are usually not sufficient to pinpoint what happens in the infrared, where the system gets strongly coupled and perturbation theory has a limited power in predicting the phase and global symmetry realization patterns.  

The tools which allow these new results come from  the idea of the generalized symmetries,  of  gauging some 1-form discrete center symmetries  and  studying the consequences  of mixed-'t Hooft-anomaly-matching conditions  \cite{GKSW}\sp\cite{Wan:2018djl}. Most concrete applications of these new techniques  so far   refer to vectorlike gauge theories, such as pure $SU(N)$ Yang-Mills,  or adjoint QCD,  where there is an exact center symmetry (${\mathbbm Z}_N$ for $SU(N)$  theories), or  QCD  where the color center symmetry can be combined with
$U(1)_V$ to give a  color-flavor locked 1-form center symmetry. 
  In these, vectorlike, gauge theories, the results from the new approach can be corroborated by the extensive literature, based on 
some general theorems \cite{Vafa:1983tf,Vafa:1984xg}, on lattice simulations \cite{Debbio:2015ila}\sp\cite{Athenodorou:2014eua}, on the effective Lagrangians \cite{Witten:1983tx}\sp\cite{Witten:1980sp},  on 't Hooft anomaly analysis  \cite{tHooft:1979rat},  on the powerful exact results in ${\cal N}=2$ supersymmetrie theories \cite{SW1,SW2},   or on   some other theoretical ideas such as the space compactification combined with semi-classical analyses \cite{unsal}\sp\cite{Yaffe}.  

Most of these theoretical tools are however unavailable for the study of strongly-coupled {\it  chiral} gauge theories, except for some general  wisdom,  the large-$N$ approximation,  and  the  't Hooft anomaly considerations.  Together, they offer significant, but not very stringent,  
information on the infrared dynamics, phases, and symmetry realization  (see \cite{Raby}\sp\cite{BK}).  Such a situation is doubtlessly limiting our capability of  utilizing chiral gauge theories 
in the context of realistic model building beyond the standard model, e.g.,  with composite fermions, with composite Higgs bosons, or  with dynamical composite models for dark matter, and so on.

 It was these considerations that recently motivated the present authors   to  apply some of the new concepts and techniques to chiral gauge theories, to see if new insights in the physics of these theories   can be gained by doing so \cite{BKL1,BKL2}.   In particular,  in  \cite{BKL2},  a simple class of  $SU(N)$ gauge theories with Weyl fermions  
\beq
   \psi^{ij}\,, \quad    \eta_i^A\, , \qquad  {\footnotesize  i,j = 1, \ldots, N\;,\quad A =1, \ldots , N+4}\;,
\eeq
in the direct-sum  representation
\be       \yng(2) \oplus   (N+4) \,{\bar   {{\yng(1)}}}\;,     \ee
 (``$\psi\eta$  model")   was studied.   For even  $N$  the  (nonanomalous) symmetry of the system is 
 \be
SU(N)\times G_{\mathrm{f}}\;, \qquad  G_{\mathrm{f}}=\frac{SU(N+4) \times  U(1)_{\psi\eta}  \times (\mathbb{Z}_2)_F } { \mathbb{Z}_{N}\times \mathbb{Z}_{N+4}}\;,
\label{symmetryNeven}
\ee
where $U(1)_{\psi\eta}$ is the anomaly-free combination of  $U(1)_{\psi}$ and $U(1)_{\eta}$,  and   $(\mathbb{Z}_2)_F$  is the fermion parity,    $\psi, \eta \to  -\psi, -\eta$. 

In spite of the presence of fermions in the fundamental representation of $SU(N)$ the system turns out to possess, classically,  an exact  discrete ${\mathbbm Z}_N$ center 1-form symmetry \footnote{Let us recall that a 1-form symmetry acts on extended operators  such as closed Wilson or Polyakov loops, but not on a local operator as in  conventional  (0-form) symmetries.}, 
\be      {\mathbbm Z}_N  =   SU(N) \cap \{  U(1)_{\psi\eta}  \times   (\mathbb{Z}_2)_F \}\;.   \label{colorflavor}
\ee
Indeed, the gauge transformation with $\rme^{\frac{2\pi\im}{N}}\in \mathbb{Z}_N\subset SU(N)$, 
\be
\psi\to \rme^{\frac{4\pi\im}{N}}\psi\;,\; \qquad \eta \to \rme^{-\frac{2\pi\im}{N}}\eta\;,    \label{ZNpsieta}
\ee
can be undone by the following $(\mathbb{Z}_2)_F \times  U(1)_{\psi\eta}  $ transformation:
\be
\psi \to (-1)\, \rme^{\im {N+4\over 2}{2\pi\over N}}\psi = \rme^{-\im {N\over 2}{2\pi\over N}}   \, \rme^{\im {N+4\over 2}{2\pi\over N}}\psi \;, \qquad \eta\to (-1)\, \rme^{-\im {N+2\over 2}{2\pi\over N}}\eta=    \rme^{\im {N\over 2}{2\pi\over N}}     \, \rme^{-\im {N+2\over 2}{2\pi\over N}}\eta\;. \label{ZNequiv}
\ee
Note that the odd elements of $\mathbb{Z}_N$ belong to the disconnected component of $U(1)_{\psi\eta}\times (\mathbb{Z}_2)_F$ while the even elements belong to the identity component.  

The central idea is now to ``gauge"  this discrete,  color-flavor locked   1-form $\mathbb{Z}_N$ symmetry. 
Remember that the unfamiliar-sounding  expression of gauging a discrete symmetry means simply that field configurations related by it are identified and the redundancy eliminated. This implies redefinition of the path-integral sum over the gauge field configurations appropriately.  By applying this to  the 1-form  ${\mathbbm Z}_N$  of an $SU(N)$ gauge theory, one arrives at an  $\frac{SU(N)}{{\mathbbm Z}_N}$ gauge system,  with consequent  $\frac{1}{N}$  fractional instanton numbers.\footnote{In \cite{BKL2} we have gauged also the 1-form center 
symmetry ${\mathbbm Z}_{N+4} \subset SU(N+4)$,   but the conclusion of the work did not depend on it.   Here and in the rest of the present work,    only the 
``color-flavor locked"  ${\mathbbm Z}_N$  center symmetry will be considered.  
   }
Concretely, this can be done by introducing the 2-form gauge fields $\big(B_\rmc^{(2)}$, $B_\rmc^{(1)}\big)$,   
\be  N  B_\rmc^{(2)} = d  B_\rmc^{(1)}\;,    \label{1fgconstraint}
\ee
and coupling to them the  $SU(N)$  gauge fields  $a$ and $U(1)_{\psi\eta}  \times   (\mathbb{Z}_2)_F$ gauge fields,  $A$ and $ A_2^{(1)}$,    appropriately.  
As for the $SU(N)$ gauge field $a$, this can be achieved by embedding it  into a  $U(N)$ gauge field $\widetilde{a}$ as 
\be
\widetilde{a}=a+{1\over N}B^{(1)}_\rmc,
\ee
and requiring the whole system to be invariant under the 1-form gauge transformation, 
\begin{align}  B_\rmc^{(2)} & \to B_\rmc^{(2)}+\diff \lambda_\rmc\;, \qquad
 B_\rmc^{(1)}  \to B_\rmc^{(1)}+ N  \lambda_\rmc    \;,   \nonumber \\ 
 \widetilde{a} &\to \widetilde{a}+\lambda_\rmc\;.
   \label{werequire} 
\end{align}
As the ${\mathbbm Z}_N$  is a color-flavor locked symmetry,  Eq.~(\ref{colorflavor}),  the   $U(1)_{\psi\eta} $ and  $  (\mathbb{Z}_2)_F$ gauge fields must also be transformed simultaneously:  
\be  A\to A-\lambda_\rmc\;, \qquad    A_2^{(1)}\to A_{2}^{(1)}+{N\over 2}\lambda_\rmc   \;. \label{simult}\;
\ee
The  relation (\ref{1fgconstraint})  indicates that one has now an  $\frac{SU(N)}{{\mathbbm Z}_N}$ connection rather than  $SU(N)$. It implies that 
there are nontrivial 't Hooft fluxes carried by the   
 gauge fields
 \be      \frac{1}{2\pi} \int_{\Sigma_2}    B_\rmc^{(2)}    =     \frac{ n_1 }{N}\;,    \qquad     n_1 \in   {\mathbbm Z}_N\;,   \label{Byconstruction2}
 \ee
  in a closed two-dimensionl subspace,    ${\Sigma_2}$.    On topologically nontrivial four dimensional spacetime
 of Euclidean signature  containing such  subspaces one has then 
 \footnote{Throughout, a   compact differential-form notation is used.  For instance, 
  $a \equiv  T^c  A_{\mu}^c(x)   \,dx^{\mu}$;    $F= d a + a^2 $\;;
$F^2 \equiv  F \wedge F =   \frac{1}{2}  F^{\mu \nu} F^{\rho \sigma} dx_{\mu} dx_{\nu} dx_{\rho} dx_{\sigma}= \frac{1}{2}
\epsilon_{\mu \nu \rho \sigma}   F^{\mu \nu} F^{\rho \sigma}  d^4x =   F^{\mu \nu} {\tilde F}_{\mu \nu}  d^4x$\,, and so on. }  
  \be      \frac{1}{8\pi^2} \int_{\Sigma_4}   (B_\rmc^{(2)})^2   =   \frac{n }{N^2}\;,\label{Byconstruction4}
 \ee
   where $n \in   {\mathbbm Z}_N$.

The fermion kinetic term with the background gauge field is obtained by the minimal coupling procedure as  \footnote{The  ${\mathbbm Z}_N$ charges of  $A$ and $ A_2^{(1)}$ in  (\ref{naive}) are determined by the way  $U(1)_{\psi\eta} $ and  $(\mathbb{Z}_2)_F$  together
reproduce    $\psi \to  e^{4\pi i / N} \psi $ and   $\eta \to  e^{- 2\pi i / N} \eta $, as the reader can easily check. See \cite{BKL2}.
 }
\bea
&&\overline{\psi}\gamma^{\mu}\Big(\partial +\calR_{\rmS}(\widetilde{a})+{N+4\over 2}A+A_2\Big)_{\mu}P_\rmL\psi + \;  \nonumber \\
&&  \overline{\eta}\gamma^{\mu}\Big(\partial +\calR_{\rmF^*}(\widetilde{a}) -{N+2\over 2}A-A_2\Big)_{\mu}P_\rmL\eta\;, \label{naive}
\eea
with the obvious notation.
We  compute the anomalies by applying the Stora-Zumino descent procedure starting with a $6D$ anomaly functional \footnote{In going from (\ref{naive}) to (\ref{tensors00}) term are arranged so that the expression inside each bracket be 1-form gauge invariant.}
\bea 
 {\cal T}_1 &=&  {1\over 24\pi^2}\, {\tr}_{\calR_\rmS}  \left[    \calR_\rmS\big(F(\tilde{a})-B^{(2)}_\rmc\big) +{N+4\over 2}\Big({\diff} A+B^{(2)}_\rmc    \Big)  +\Big(\diff A_2^{(1)}-{N\over 2}B^{(2)}_\rmc   \Big)
   \right]^3  \;,     \nonumber \\
   {\cal T}_2  &=&  {1\over 24\pi^2}\, {\tr}_{\calR_\rmF^*} \left[ -\big(F(\widetilde{a})-B^{(2)}_\rmc\big)   -{N+2\over 2} \Big({\diff} A+B^{(2)}_\rmc  \Big)   -\Big(\diff A_2^{(1)}-{N\over 2}B^{(2)}_\rmc   \Big)
   \right]^3   \;.    \nonumber \\
 \label{tensors00} 
   \eea  
   The rest of the  procedure     for computing the  $(\mathbb{Z}_2)_F$ anomaly is standard:  (i)  one first integrates to get the $5D$ boundary action containing $A_2^{(1)}$
   (WZW action);  (ii)  the variations  of the form 
   \be    \delta A_2^{(1)}  =   \frac{1}{2}   \de \delta A_2^{(0)}\;, \qquad    \delta A_2^{(0)} = \pm 2\pi\;,      \label{usual1}
   \ee 
   leads to,  via  the anomaly-in-flow,  the seeked-for  anomaly in the $4D$  theory.     The result is 
   \be  \delta S =        -  N^2  \,      {1\over 8 \pi^2}  \int_{\Sigma^4} (B^{(2)}_\rmc)^2    \,   \frac{1}{2}  \delta A_2^{(0)}  = - N^2 \times \frac{\mathbbm Z}{N^2} \,  ({\pm \pi})
=\pm    \pi   \times  {\mathbbm Z}\;:     \label{usual2}
\ee
   the partition function changes sign, under     $\psi, \eta \to  -\psi, -\eta$,  that is,     there is a  $(\mathbb{Z}_2)_F$ anomaly. 
   
As  the $({\mathbbm Z}_{2})_F  - [{\mathbbm Z}_{N}]^2$  mixed anomaly   is obviously absent in the IR,   we conclude that  the confining chirally symmetric vacuum, in which conventional 't Hooft anomalies are saturated in the infrared 
 by  massless composite ``baryons" 
   \be     {\cal B}^{AB}=    \psi^{ij}  \eta_i^A  \eta_j^B \;,\qquad  A,B=1, \ldots, N+4\;,\label{baryons10}
\ee 
(antisymmetric in  $A \leftrightarrow B$), is not the correct vacuum of the system.   As shown in \cite{BKL2}, the dynamical Higgs vacuum, characterized by bifermion condensates,  
\be    \brc  \psi^{ij}   \eta_i^B \ckt =\,   c \,  \Lambda^3   \delta^{j B}\ne 0\;,   \qquad   j, B=1, \dots  N\;,   \qquad c \sim O(1) \label{cflocking}
\ee
is instead  found to be  fully consistent.   

Several subtle features of the calculation and in the interpretation of the results are discussed carefully in \cite{BKL2}.

The purpose of the present work is to investigate if  the result found in the $\psi \eta$ model   extends naturally to a wider class of the  so-called  Bars-Yankielowicz and the generalized Georgi-Glashow models. 
The gauge group is taken to be $SU(N)$, and the matter fermion content is   ($p$ is a natural number)
\be       \yng(2) \oplus   (N+4+p) \,{\bar   {{\yng(1)}}}\;\oplus   p \,{   {{\yng(1)}}}\; 
\ee
for the former  (let us call them $\{\S,N,p\}$ models),  and 
\be       \yng(1,1) \oplus   (N-4+p) \,{\bar   {{\yng(1)}}}\;\oplus   p \,{   {{\yng(1)}}}\; 
\ee
for the latter  ($\{\A,N,p\}$ models).   We will find that for all $N$ and $p$,  both even, the system possesses a $(\mathbb{Z}_2)_F$ symmetry,  which is nonanomalous, i.e., respected by standard instantons.   Also, these models all enjoy a   ``color-flavor locked"  ${\mathbbm Z}_N$  center symmetry,  in spite of the presence of fermions in the fundamental (or anti-fundamental) representation.
It is thus possible to gauge this center symmetry  and study if, by doing so, the $(\mathbb{Z}_2)_F$ symmetry becomes anomalous,    as happened 
in the $\{\S,N, 0\}$ model.

The paper is organized as follows.  In Sec.~\ref{models}    we discuss the conventional 't Hooft-anomaly-matching analysis in all these models. A good part of this section is a review of \cite{Raby}-\cite{ADS}, but there are some new results, especially concerning the Higgs phase, which we need later. As the global symmetry group is relatively large,  the fact that one can find a set of gauge-invariant composite fermions which satisfy all the anomaly-matching equations  at all,  assuming the system to confine,  is quite remarkable.  Also, in all these models we find an alternative phase, also consistent with the anomaly matching criterion,  characterized by certain  bifermion condensates breaking color dynamically (dynamical Higgs phase)  accompanied by a partial breaking of the global symmetry.  

In the conventional 't Hooft anomaly matching equations, only the perturbative (local) aspect of the flavor symmetry group matters, though nonperturbative (instanton) effects of the strong $SU(N)$ gauge interactions are taken into account.   In  Sec.~\ref{symmetries},  the symmetry of these models is re-analyzed more carefully, taking into account the global properties (e.g., the connectedness).  

In Sec.~\ref{sec:mixedS}  we calculate and find a mixed anomaly of the 
type,  $({\mathbbm Z}_{2})_F  - [{\mathbbm Z}_{N}]^2$,  in all
 models with $N$ and $p$ both even,  whereas such an anomaly is absent in the infrared (IR) in a confining vacuum with full global symmetry - one of the candidate vacua 
allowed by the conventional anomaly matching argument.  Consistency implies that these vacua cannot be realized dynamically in the infrared, in all 
$\{\S,N,p\}$  and $\{\A,N,p\}$ models, with  $N$ and $p$ are both even.

We summarize and  discuss our results in Sec.~\ref{conclude}.

\section{Theories and possible phases
 \label{models}}

\subsection{$\{\S,N,p\}$ models}

The first class of theories is the  $\psi \eta$ model with additional $p$ pairs of fundamental and anti-fundamental fermions.   Namely,  the  model   is  an $SU(N)$ gauge theory with Weyl fermions 
\beq
   \psi^{ij}\,, \quad    \eta_i^A\, \,, \quad    \xi^{i,a}  
\eeq
in the direct-sum  representation
\be       \yng(2) \oplus   (N+4+p) \,{\bar   {{\yng(1)}}}\;\oplus   p \,{   {{\yng(1)}}}\; .
\ee
The indices run as
\beq
\footnotesize  i,j = 1,\ldots, N\;,\quad A =1,\ldots , N+4+p  \;,\quad a =1,\ldots , p  \;.
\eeq
These theories (the Bars-Yankielowicz models) will be denoted as $\{\S,N,p\}$ below. The  $\psi \eta$ model corresponds to  $\{\S,N,0\}$.
The first coefficient of the beta function is
\be   b_0 = 11N -  (N+2) - (N+4 + 2p) = 9N-6-2p\;,  \label{betafn}
\ee 
and $p$ is limited by $\frac{9}{2}N-3$ before asymptotic freedom (AF) is lost. In the limit $N$ fixed and $p \to \infty$ we recover ordinary QCD with $p$ flavors, although this is outside the regime of AF.
The classical symmetry group is
  \be  SU(N)_{\rmc} \times  U(1)_{\psi} \times  U(N+4+p)_{\eta} \times  U(p)_{\xi}\;.  \label{groups}
    \ee
 We discuss for the moment only $0$-form symmetries, leaving a more detailed discussion of the symmetry group to   Sec.~\ref{symmetries}.\footnote{To be precise, (\ref{groups}) is a covering space of the ``real'' symmetry group.  As the conventional 't Hooft analysis depends only on the algebra of the symmetry group, this 
 is for the moment sufficient. }
Anomaly breaks the symmetry group (\ref{groups}) to
  \bea
&&p=0:   \quad  SU(N)_{\rmc}  \times   SU(N+4)_{\eta}   \times  U(1)_{\psi\eta}\;,  \nn \\
&&p=1:    \quad  SU(N)_{\rmc}  \times   SU(N+5)_{\eta}  \times   U(1)_{\psi\eta}\times  U(1)_{\psi\xi}\;,  
\nn \\
&&p>1:   \quad  SU(N)_{\rmc}  \times   SU(N+4+p)_{\eta}  \times  SU(p)_{\xi}  \times  U(1)_{\psi\eta}\times  U(1)_{\psi\xi}\;,  
 \label{groupsb}
    \eea
    where   the anomaly-free combination of  $U(1)_{\psi}$ and $U(1)_{\eta}$ is
    \be
U(1)_{\psi\eta} : \qquad  \psi\to \rme^{\im (N+4+p)\alpha}\psi\;, \quad  \eta \to \rme^{-\im (N+2)\alpha}\eta\;, \label{upe}
\ee  
 with $\alpha \in \mathbbm{R}$,  and   the anomaly-free combination of  $U(1)_{\psi}$ and $U(1)_{\xi}$ is
    \be
U(1)_{\psi\xi} : \qquad \psi\to \rme^{\im p \beta}\psi\;, \quad  \xi \to \rme^{-\im (N+2)\beta}\xi\;, \label{upx}
\ee  
with $\beta \in \mathbbm{R}$.
The choice of the two unbroken $U(1)$'s is somehow arbritrary, for example also $U(1)_{\eta\xi} $ 
\be
U(1)_{\eta\xi} : \qquad  \eta \to \rme^{\im p \gamma}\eta \;, \quad  \xi \to \rme^{-\im (N+4+p)\gamma}\xi\;,
\label{upex}
\ee  
with $\gamma \in \mathbbm{R}$ could be chosen as a generator. In Table~\ref{suv} we summarize the fields and how they transform under the symmetry group.
\begin{table}[h!t]
  \centering 
  \small{\begin{tabular}{|c|c|c|c|c|c|  }
\hline
\su      &  $SU(N)_{\rm c}  $    &  $ SU(N+4+p)$    &  $ SU(p)$     &   $ {U}(1)_{\psi\eta}   $  &   $ {U}(1)_{\psi\xi}   $  \\
\hline 
\sbu  $\psi$   &   $ { \yng(2)} $  &    $  \frac{N(N+1)}{2} \cdot (\cdot) $    & $   \frac{N(N+1)}{2} \cdot (\cdot)  $  & $N +4 +p$ & $p $  \\
  $ \eta$      &   $  (N+4+p)  \cdot   {\bar  {\yng(1)}}   $     & $N  \cdot  {\yng(1)}  $     & $N (N+4+p) \, \cdot  (\cdot)   $   &$-(N+2)$&$0$\\ 
$ \xi$      &   $  p \cdot   {  {\yng(1)}}   $     & $N p \, \cdot  (\cdot)   $     &  $ N  \cdot   { {\yng(1)}} $ &$0$&$-(N+2)$ \\
\hline   
\end{tabular}}
  \caption{\footnotesize The multiplicity, charges and the representation are shown for each  set of fermions in the $\{\S,N,p\}$ model.  $(\cdot)$ stands for a singlet representation.
}\label{suv}
\end{table}
There are also discrete  unbroken symmetries of the three $U(1)$'s:   $({\mathbbm Z}_{N+2})_{\psi}$, $({\mathbbm Z}_{N+4+p})_{\eta}$ and $({\mathbbm Z}_{p})_{\xi}$. The relation between these  discrete symmetries and  the continuous non-anomalos group $U(1)_{\psi\eta} \times
U(1)_{\psi\xi} $ will be discussed in Sec.~\ref{symmetries}.

\subsection{$\{\A,N,p\}$ models}

The second class of models we are interested  are  $SU(N)$ gauge theories with Weyl fermions 
\beq
   \chi^{ij}\,, \quad    \eta_i^A\, \,, \quad    \xi^{i,a}  
\eeq
in the direct-sum  representation
\be       \yng(1,1) \oplus   (N-4+p) \,{\bar   {{\yng(1)}}}\;\oplus   p \,{   {{\yng(1)}}}\;. 
\ee
The indices run as
\beq
\footnotesize  i,j = 1,\ldots, N\;,\quad A =1,\ldots , N-4+p  \;,\quad a =1,\ldots , p  \;.
\eeq
These (the generalized Georgi-Glashow) models will be indicated as $\{\A,N,p\}$.
The first coefficient of the beta function is
\be   b_0 = 11N -  (N-2) - (N-4 +2p) = 9N+6-2p\;.  \label{betafn}
\ee
Here  $p$ will be assumed to be less than $\frac{9}{2}N+3$ so as to  maintain AF. 
The symmetry group is
  \be  SU(N)_{\rmc} \times  U(1)_{\chi} \times  U(N-4+p)_{\eta} \times  U(p)_{\xi}\;.  \label{groupa}
    \ee
Anomaly breaks this group  to
  \bea
&&p=0:   \quad  SU(N)_{\rmc}  \times   SU(N-4)_{\eta}   \times  U(1)_{\chi\eta}\;,  \nn \\
&&p=1:    \quad  SU(N)_{\rmc}  \times   SU(N-3)_{\eta}  \times   U(1)_{\chi\eta}\times  U(1)_{\chi\xi}\;,  
\nn \\
&&p>1:   \quad  SU(N)_{\rmc}  \times   SU(N-4+p)_{\eta}  \times  SU(p)_{\xi}  \times  U(1)_{\chi\eta}\times  U(1)_{\chi\xi}\;,  
 \label{groupab}
    \eea
    where the anomaly-free combination of  $U(1)_{\chi}$ and $U(1)_{\eta}$ is
    \be
U(1)_{\chi\eta} : \  \chi\to \rme^{\im (N-4+p)\alpha}\chi\;, \qquad  \eta \to \rme^{-\im (N-2)\alpha}\eta\;, 
\label{uce}
\ee  
  and   the anomaly-free combination of  $U(1)_{\psi}$ and $U(1)_{\xi}$ is
    \be
U(1)_{\chi\xi} : \ \chi\to \rme^{\im p \beta}\chi\;,   \qquad  \xi \to \rme^{-\im (N-2)\beta}\xi\;.
\label{ucx}
\ee  
Another possible anomaly-free combination is  $U(1)_{\eta\xi} $:
\be
U(1)_{\eta\xi} : \ \eta \to \rme^{\im p \gamma}\eta \;, \qquad  \xi \to \rme^{-\im (N-4+p)\gamma}\xi\;.
\label{ucex}
\ee 
In Table~\ref{sua} we summarize the fields and how they transform under the symmetry group.
\begin{table}[h!t]
  \centering 
  \small{\begin{tabular}{|c|c|c|c|c|c|  }
\hline
\su      &  $SU(N)_{\rm c}  $    &  $ SU(N-4+p)$    &  $ SU(p)$     &   $ {U}(1)_{\chi\eta}   $  &   $ {U}(1)_{\chi\xi}   $  \\
\hline 
\sbuu   $\chi$   &   $ { \yng(1,1)} $  &    $  \frac{N(N-1)}{2} \cdot (\cdot) $    & $   \frac{N(N-1)}{2} \cdot (\cdot)  $  & $N -4 +p$ & $p $  \\
  $ \eta$      &   $  (N-4+p)  \cdot   {\bar  {\yng(1)}}   $     & $N  \cdot  {\yng(1)}  $     & $N (N-4+p) \, \cdot  (\cdot)   $   &$-(N-2)$&$0$\\
 
$ \xi$      &   $  p \cdot   {  {\yng(1)}}   $     & $N p \, \cdot  (\cdot)   $     &  $ N  \cdot   { {\yng(1)}} $ &$0$&$-(N-2)$ \\
\hline   
\end{tabular}}
  \caption{\footnotesize The multiplicity, charges and the representation are shown for each  set of fermions in the $\{\A,N,p\}$ model. 
}\label{sua}
\end{table}
There are also discrete  unbroken symmetries:   $({\mathbbm Z}_{N-2})_{\psi}$, $({\mathbbm Z}_{N-4+p})_{\eta}$ and $({\mathbbm Z}_{p})_{\xi}$.

\subsection{Confining phase with unbroken global symmetries   \label{unbroken}}

The standard 't Hooft anomaly matching conditions were found to   allow a chirally symmetric, confining vacuum in the model first proposed in \cite{BY}.  
    Let us assume that no condensates form, the system confines, and the flavor symmetry is unbroken. 
    
\subsubsection{$\{\S,N,p\}$  models}

   The candidate massless composite fermions  for the  $\{\S,N,p\}$  models are the left-handed gauge-invariant fields:
      \be    
 {({\cal B}_{1})}^{[AB]}=    \psi^{ij}   \eta_i^{A}  \eta_j^{B}\;,
\qquad {({\cal B}_{2})}^{a}_{A}=    \bar{\psi}_{ij}  \bar{\eta}^{i}_{A}  \xi^{j,a} \;,
\qquad {({\cal B}_{3})}_{\{ab\}}=    \psi^{ij}  \bxi_{i,a}  \bxi_{j,b}  \;,
\label{baryons10}
\ee
the first is anti-symmetric in $A \leftrightarrow B$ and the third is  symmetric in $a \leftrightarrow b$; their charges are given in Table  \ref{sir}.
Writing explicitly also the spin indices they are 
 \bea    
& {({\cal B}_{1})}^{AB, \alpha}=   \frac{1}{2} \epsilon_{\beta \gamma} \psi^{ij, \beta}  \eta_i^{A, \gamma}  \eta_j^{B, \alpha}  +  \frac{1}{2} \epsilon_{\beta \gamma} \psi^{ij, \beta}  \eta_i^{A, \alpha}  \eta_j^{B, \gamma}   
\;, &\nn \\
& {({\cal B}_{2})}^{a, \alpha}_{A}=     \epsilon_{\dot{\alpha}\dot{\beta} }  \bar{\psi}_{ij}^{\dot{\alpha}}  \bar{\eta}^{i,\dot{\beta}}_{A}  \xi^{j,a,\alpha} \;,
\qquad {({\cal B}_{3})}_{ab}^{\alpha}=    \epsilon_{\dot{\beta} \dot{\gamma}} \psi^{ij, \alpha}  \bxi_{i,a}^{\dot{\beta}}  \bxi_{j,b}^{\dot{\gamma}} 
\;: &
\label{b}
\eea
all  transforming under the  $\{\tfrac{1}{2}, 0\}$ representation of the Lorentz group.
\begin{table}[h!t]
  \centering 
  \small{\begin{tabular}{|c|c|c |c|c| c|c| }
\hline
\su      &  $SU(N)_{\rm c}  $    &  $ SU(N+4+p)$    &  $ SU(p)$     &   $ {U}(1)_{\psi\eta}   $  &   $ {U}(1)_{\psi\xi}   $  \\
  \hline     
 \sbuu     $ {{\cal B}_{1}}$      &  $   \frac{(N+4+p)(N+3+p)}{2} \cdot ( \cdot )    $         &  $ {\yng(1,1)}$        &    $  \frac{(N+4+p)(N+3+p)}{2} \cdot ( \cdot )     $   &  $-N+p $ & $p$ \\
  \hline     
  \sbbu   $ {{\cal B}_{2}}$   &    $ (N+4+p) p\cdot ( \cdot )$     &       $  p\cdot   \bar{\yng(1)}$   &     $ (N+4+p)  \cdot {\yng(1)}$      & $-(p+2)$ & $-(N+p+2)$ \\
  \hline     
   \sbbu  ${{\cal B}_{3}}$     &  $ \frac{p (p+1)}{2}  \cdot ( \cdot )   $    &    $ \frac{p (p+1)}{2}  \cdot ( \cdot )   $       &    $ \bar{\yng(2)}$       & $N+4+p$ & $2N +4 + p$\\
\hline
\end{tabular}}
  \caption{\footnotesize  Chirally symmetric phase of the  $\{\S,N,p\}$  model. 
}\label{sir}
\end{table}
Table \ref{suvsir} summarizes the anomaly matching checks, via comparison between Table~\ref{suv} and Table~\ref{sir}.
\begin{table}[h!t]
\centering
 \tiny{\begin{tabular}{|c|c|c|}
\hline
\su      &  UV    &  IR  \\
\hline 
\su $ SU(N+4+p)^3$     &   $ N $     &  $ N+p -p $ \\
\su $ SU(p)^3$     &   $ N $     &  $ N+4+p -(p+4) $ \\
\su $ SU(N+4+p)^2 - {U}(1)_{\psi\eta}$     &   $ - N(N+2)  $     &  $ -(N+2+p)(N-p) -p(p+2) $ \\
\su $ SU(N+4+p)^2- {U}(1)_{\psi\xi}$     &    $ 0 $     &  $  (N+2+p)p -p(N+p+2)  $ \\
\su $ SU(p)^2- {U}(1)_{\psi\eta}$     &   $ 0 $     &  $ -(N+4+p)(p+2) +(p+2)(N+p+4)  $ \\
\su $ SU(p)^2- {U}(1)_{\psi\xi}$     &   $ - N(N+2)   $     &  $   -(N+4+p)(N+p+2) +(p+2)(2N+p+4)$ \\
\su $ {U}(1)_{\psi\eta}^3$     &   $ \frac{N(N+1)}{2}(N+4+p)^3 - N(N+4+p) (N+2)^3 $     &  $ - \frac{(N+4+p)(N+3+p)}{2} (N-p)^3 -(N+4+p) p (p+2)^3 + $ \\\
\su      &    
&  $ +  \frac{p (p+1)}{2}(N+4+p)^3$ \\
\su $ {U}(1)_{\psi\xi}^3$     &   $ \frac{N(N+1)}{2}p^3 - N p (N+2)^3  $     &  $    \frac{(N+4+p)(N+3+p)}{2} p^3 -(N+4+p) p (N+p+2)^3  +$ \\
\su    &     
 &  $    \  +  \frac{p (p+1)}{2}(2N+4+p)^3$ \\
\su $ {\rm Grav}^2-{U}(1)_{\psi\eta} $     &  $ \frac{N(N+1)}{2}(N+4+p)  - N(N+4+p) (N+2)  $     &  $ - \frac{(N+4+p)(N+3+p)}{2} (N-p)  -(N+4+p) p (p+2) + $ \\
\su      &   
&  $ +  \frac{p (p+1)}{2}(N+4+p) $ \\
\su $ {\rm Grav}^2-{U}(1)_{\psi\xi}$       &   $\frac{N(N+1)}{2}p - N p (N+2)  $     &  $    \frac{(N+4+p)(N+3+p)}{2} p -(N+4+p) p (N+p+2)  +$ \\
\su    &     
&  $    \  +  \frac{p (p+1)}{2}(2N+4+p)$ \\
\su $ SU(N+4+p)^2-({\mathbbm Z}_{N+2})_{\psi}  $     &  $0$    & $N+2+p-p = 0 \ {\rm mod} \ N+2$    \\
\su $ SU(p)^2- ({\mathbbm Z}_{N+2})_{\psi} $     &   $0$    & $-(N+4+p) + p+2  = 0 \ {\rm mod} \ N+2 $     \\
\su $ {\rm Grav}^2-({\mathbbm Z}_{N+2})_{\psi}  $     &  $1$      & $1-1+1$  \\
\hline   
\end{tabular}}
  \caption{\footnotesize  Anomaly matching checks for the IR chiral symmetric phase of the $\{\S,N,p\}$  model.    For $N$ odd, the last three equalities  
  are consequences of other equations.
}\label{suvsir}
\end{table}

\subsubsection{$\{\A,N,p\}$  models}

The candidate massless composite fermions  for the  $\{\A,N,p\}$  model are:
  \be    
 {({\cal B}_{1})}^{\{AB\}}=    \chi^{ij}  \eta_i^{A}  \eta_j^{B} \;,
\qquad {({\cal B}_{2})}^{a}_{A}=    \bar{\chi}_{ij}  \bar{\eta}^{i}_{A}  \xi^{j,a} \;,
\qquad {({\cal B}_{3})}_{[ab]}=    \chi^{ij}  \bxi_{i,a}  \bxi_{j,b} \;,
\label{baryons20}
\ee
the first  symmetric in $A \leftrightarrow B$ and the third anti-symmetric in $a \leftrightarrow b$.
Writing the spin indices explicitly  they are: 
 \bea   
 & {({\cal B}_{1})}^{AB, \alpha}=  \frac{1}{2}   \epsilon_{\beta\gamma}  \, \chi^{ij, \beta}  \eta_i^{A, \gamma}  \eta_j^{B, \alpha}  +
  \frac{1}{2}   \epsilon_{\beta\gamma}  \, \chi^{ij, \beta}  \eta_i^{A, \alpha}  \eta_j^{B, \gamma}    \;,    
\nn \\    &{({\cal B}_{2})}^{a, \alpha}_{A}=     \epsilon_{\dot{\beta} \dot{\gamma}} \   \bar{\chi}_{ij}^{\dot \beta}     \bar{\eta}^{i, \dot{\gamma}}_{A}  \xi^{j,a, \alpha} \;,  \qquad    {({\cal B}_{3})}_{ab}=   \epsilon_{\dot{\beta} \dot{\gamma}} \       \chi^{ij}  \bxi_{i,a}^{\dot \beta}  \bxi_{j,b}^{\dot{\gamma}}  
\;. &
\label{baryons1}
\eea
All anomaly triangles  are saturated  by these candidate massless composite fermions,  see Table~\ref{suvsirBis}  (Table~\ref{air} vs Tab.~\ref{sua}). 
\begin{table}[h!t]
  \centering 
  \small{\begin{tabular}{|c|c|c |c|c| c|c| }
\hline
\su      &  $SU(N)_{\rm c}  $    &  $ SU(N-4+p)$    &  $ SU(p)$     &   $ {U}(1)_{\chi\eta}   $  &   $ {U}(1)_{\chi\xi}   $  \\
  \hline     
 \sbbu      $ {\cal B}_{1}$   &  $  \frac{(N-4+p)(N-3+p)}{2} \cdot ( \cdot )    $         &  $ {\yng(2)}$        &    $  \frac{(N-4+p)(N-3+p)}{2} \cdot ( \cdot )    $   &  $-N+p $ & $p$ \\
  \hline     
 \sbbu    $ {\cal B}_{2}$  &   $ (N-4+p) p \cdot ( \cdot )  $   &        $ p \cdot  \bar{\yng(1)}$   &     $  (N-4+p) \cdot  {\yng(1)}$     & $-(p-2)$ & $-(N+p-2)$\\
  \hline     
   \sbbuu  $ {\cal B}_{3}$   &    $  \frac{p (p-1)}{2}  \cdot ( \cdot )  $  &      $ \frac{p (p-1)}{2}  \cdot ( \cdot )  $ &    $ \bar{\yng(1,1)}$   & $N-4+p$ & $2N -4 +p$\\
\hline
\end{tabular}}
  \caption{\footnotesize  IR massless fermions in the chirally symmetric phase of the  $\{\A,N,p\}$  model. 
}\label{air}
\end{table}
\begin{table}[h!t]
\centering
 \tiny{\begin{tabular}{|c|c|c|}
\hline
\su      &  UV    &  IR  \\
\hline 
\su $ SU(N-4+p)^3$     &   $ N $     &  $ N+p -p $ \\
\su $ SU(p)^3$     &   $ N $     &  $ N-4+p -(p-4) $ \\
\su $ SU(N-4+p)^2 - {U}(1)_{\chi\eta}$     &   $ - N(N-2)  $     &  $ -(N-2+p)(N-p) -p(p-2) $ \\
\su $ SU(N-4+p)^2- {U}(1)_{\chi\xi}$     &    $ 0 $     &  $  (N-2+p)p -p(N+p-2)  $ \\
\su $ SU(p)^2- {U}(1)_{\chi\eta}$     &   $ 0 $     &  $ -(N-4+p)(p-2) +(p-2)(N-4+p)  $ \\
\su $ SU(p)^2- {U}(1)_{\chi\xi}$     &   $ - N(N-2)   $     &  $   -(N-4+p)(N+p-2) +(p-2)(2N-4+p +0)$ \\
\su $ {U}(1)_{\chi\eta}^3$     &   $ \frac{N(N-1)}{2}(N-4+p)^3 - N(N-4+p) (N-2)^3  $     &  $ - \frac{(N-4+p)(N-3+p)}{2} (N-p)^3 -(N-4+p) p (p-2)^3 + $ \\\
\su      &        &  $ +  \frac{p (p-1)}{2}(N-4+p)^3$ \\
\su $ {U}(1)_{\chi\xi}^3$     &   $ \frac{N(N-1)}{2}p^3 - N p (N-2)^3  $     &  $    \frac{(N-4+p)(N-3+p)}{2} p^3 -(N-4+p) p (N+p-2)^3  +$ \\
\su    &      &  $    \  +  \frac{p (p-1)}{2}(2N-4+p)^3$ \\
\su $ {\rm Grav}^2-{U}(1)_{\chi\eta} $     &  $ \frac{N(N-1)}{2}(N-4+p)  - N(N-4+p) (N-2)   $     &  $ - \frac{(N-4+p)(N-3+p)}{2} (N-p)  -(N-4+p) p (p-2) + $ \\
\su      &       &  $ +  \frac{p (p-1)}{2}(N-4+p) $ \\
\su $ {\rm Grav}^2-{U}(1)_{\chi\xi}$       &   $ \frac{N(N-1)}{2}p - N p (N-2) $     &  $    \frac{(N-4+p)(N-3+p)}{2} p -(N-4+p) p (N+p-2)  +$ \\
\su    &      &  $    \  +  \frac{p (p-1)}{2}(2N-4+p)$ \\
\su $ SU(N-4+p)^2-({\mathbbm Z}_{N-2})_{\chi}  $     &  $0$    & $N-2+p-p = 0 \ {\rm mod} \ N-2$    \\
\su $ SU(p)^2- ({\mathbbm Z}_{N-2})_{\chi} $     &   $0$    & $-(N-4+p) + p-2  = 0 \ {\rm mod} \ N-2 $     \\
\su $ {\rm Grav}^2-({\mathbbm Z}_{N-2})_{\chi}  $     &  $1$      & $1-1+1$  \\
\hline   
\end{tabular}}
  \caption{\footnotesize  Anomaly matching checks for the IR chiral symmetric phase of the $\{\A,N,p\}$  model.
}\label{suvsirBis}
\end{table}

\subsection{Dynamical Higgs phase in the $\{\S,N,p\}$ models   \label{brokenS}}

The broken phase for the $\{\S,N,0\}$, $\psi \eta$ model has also been studied earlier \cite{ADS,BKS}. The composite scalar $\psi \eta$ in the maximal attractive channel is in the fundamental of both the gauge group and the flavor group.  All details can be found in the references.

Something interesting happens for $p > 0$. Now there is another channel, $  \xi \eta $, which is gauge invariant and charged under the flavor group. We thus have a competition between two possible symmetry breaking channels, $\psi \eta$ and $  \xi \eta $. 
We assume that both  condensates occur in the following way:
\bea
&&  \brc  \psi^{ij}   \eta_i^B \ckt =\,   c_{\psi\eta} \,  \Lambda^3   \delta^{j B}\ne 0\;,   \qquad   j,B=1,\dots,  N\;,     \nonumber \\
&& \brc  \xi^{i,a}   \eta_i^A \ckt =\,   c_{\eta\xi} \,  \Lambda^3   \delta^{aA}\ne 0\;,   \qquad  a = 1,\dots, N\;,  \quad  A=N+1,\dots, N+ p \;, 
  \eea
where $\Lambda$ is the renormailization-invariant scale dynamically generated by the gauge interactions and $ c_{\eta\xi} ,  c_{\psi\eta} $ are coefficients both of order one. 
According to the tumbling scenario \cite{Raby}, the first condensate to occur is in  the maximally attractive channel (MAC).  The strengths of the one-gluon exchange potential for the two channels
 \bea   &   \psi \Big(\raisebox{-2pt}{\yng(2)}\Big) \,  \eta \Big(\bar{\raisebox{-2pt}{\yng(1)}}\Big)      \qquad  \  & {\rm forming}    \qquad      \raisebox{-2pt}{\yng(1)}\;,  \nonumber    \\
&  \quad  \   \xi \Big({\raisebox{-2pt}{\yng(1)}}\Big) \, \eta \Big(\bar{\raisebox{-2pt}{\yng(1)}}\Big)    \qquad   \ & {\rm forming}  \qquad  (\cdot)  \;, 
 \eea
are, respectively,
 \bea      \frac{N^2-1}{2N}  -      \frac{ (N+2)(N-1)}{N}  -   \frac{N^2-1}{2N}  &=&  -      \frac{ (N+2)(N-1)}{N} \;,\nonumber \\
    0 -   2      \frac{N^2-1}{2N}    &=&  -       \frac{N^2-1}{N}  \;.
 \eea
So the  $ \psi \eta$ channel  is slightly more attractive, but such a perturbative argument is not really significant and we assume here that both types of condensates are formed.     

The resulting pattern of symmetry breaking is
\bea
&& SU(N)_{\rmc}  \times   SU(N+4+p)_{\eta}  \times  SU(p)_{\xi}  \times  U(1)_{\psi\eta}\times  U(1)_{\psi\xi} \nn \\
&&  \xrightarrow{\brc  \xi   \eta \ckt , \brc  \psi \eta \ckt}      SU(N)_{{\rm cf}_{\eta}}  \times   SU(4)_{\eta}  \times  SU(p)_{\eta\xi}  \times  U(1)_{\psi \eta}^{\prime} \times  U(1)_{\psi \xi}^{\prime} \;.
\label{symbres}
\eea
At the end the color gauge symmetry is completely (dynamically) broken, leaving  color-flavor diagonal  $SU(N)_{{\rm cf}_{\eta}} $ symmetry.
$U(1)_{\psi \eta}^{\prime}$ and $U(1)_{\psi \xi}^{\prime}$ are  combinations respectively  of $U(1)_{\psi \eta}$ (\ref{upe}) and $U(1)_{\xi \eta}$ (\ref{upx}) with  the element of $ SU(N+4+p)_{\eta}$ generated by 
\be   t_{SU(N+4+p)_{\eta}}= \left(\begin{array}{c|c|c}
(-\alpha (p+2) - p\beta) {\bf 1}_{N\times N}&&\\
\hline
&\frac{\alpha(N-p) - \beta p}{2}{\bf 1}_{4\times 4}&\\
\hline
&&(\alpha+\beta) (N+2) {\bf 1}_{p\times p}\\
\end{array}\right) \;.
\ee
Making the decomposition of the fields in the direct sum of representations in  the subgroup one gets Table~\ref{brsuv}.
\begin{table}[h!t]
{  \centering 
  \small{\begin{tabular}{|c|c |c|c|c|c|  }
\hline
\su  &   $SU(N)_{{\rm cf}_{\eta}}   $    &  $SU(4)_{\eta}$     &  $ SU(p)_{\eta\xi}$ &   $  U(1)_{\psi \eta}^{\prime}$   &   $ U(1)_{\psi \xi}^{\prime}$ \\
 \hline
  \sbu $\psi$   &   $ { \yng(2)} $  &    $  \frac{N(N+1)}{2} \cdot   (\cdot) $    & $ \frac{N(N+1)}{2} \cdot   (\cdot)   $   &  $N +4 +p$ &  $p $  \\
   $ \eta_1$      &   $  {\bar  {\yng(2)}} \oplus {\bar  {\yng(1,1)}}  $     & $N^2  \cdot  (\cdot )  $     &   $ N^2  \cdot  (\cdot ) $    &$-(N +4 +p)$ & $-p $\\
    $ \eta_2$      &   $ 4  \cdot   {\bar  {\yng(1)}}   $     & $N  \cdot  {\yng(1)}  $     &   $ 4 N  \cdot  (\cdot ) $   & $-\frac{N +4+p}{2}$  & $-\frac{p}{2}$\\
  $ \eta_3$      &   $ p  \cdot   {\bar  {\yng(1)}}$     & $ N p  \cdot  (\cdot )  $     &   $N  \cdot  \bar{\yng(1)}  $    &$0$ &$N+2$ \\
    $ \xi$      &   $ p  \cdot   { {\yng(1)}}   $     & $N p  \cdot  (\cdot )  $     &   $ N  \cdot   { {\yng(1)}}   $   & $0$  & $-(N+2)$\\
\hline 
\end{tabular}}  
  \caption{\footnotesize   UV fieds in the $\{\S,N,p\}$ model, decomposed as a direct sum of the representations of the unbroken group of  Eq.~(\ref{symbres}). }
\label{brsuv}
}
\end{table}
\begin{table}[h!t]
{  \centering 
  \small{\begin{tabular}{|c|c |c|c|c|c|  }
\hline
 \su   &  $SU(N)_{{\rm cf}_{\eta}}   $    &  $SU(4)_{\eta}$     &  $ SU(p)_{\eta\xi}$ &   $ U(1)_{\psi \eta}^{\prime}  $   &   $ U(1)_{\psi \xi}^{\prime}$ \\
 \hline
      \sbbuu $ {\cal B}_{1} $      &  $ {\bar  {\yng(1,1)}}   $         &  $  \frac{N(N-1)}{2} \cdot  (\cdot) $        &    $  \frac{N(N-1)}{2} \cdot  (\cdot) $     & $-(N +4 +p)$  &  $ -p $\\
   $ {\cal B}_{2}  $      &  $   4 \cdot {\bar  {\yng(1)}}   $         &  $N  \cdot  {\yng(1)}  $        &    $4 N  \cdot  (\cdot ) $   & $-\frac{N + 4+p}{2}$  & $-\frac{p}{2}$\\
\hline
\end{tabular}}  
  \caption{\footnotesize    IR fieds in the $\{\S,N,p\}$ model, the massless subset of the baryons in Tab.~\ref{sir} in the Higgs phase. }
\label{brsir}
}
\end{table}
The composite massless baryons are subset of  those in  (\ref{baryons10}):
\bea
 & {\cal B}_{1}^{[AB]} =  \psi^{ij}   \eta_{i}^{A}  \eta_{j}^{B} \;,  \qquad 
 {\cal B}_{2}^{[AC]} =  \psi^{ij}   \eta_{i}^{A}  \eta_{j}^{C}\;, & \nn \\
& A,B = 1, \dots, N \;,  \quad C=N+1, \dots, N+4 \;. &
\eea
It is quite straightforward (and actually almost trivial)  to verify  -  we leave it to the reader as an exercise -  that the UV-IR  anomaly matching continues to work, with  the UV fermions in Table~\ref{brsuv}
and the IR fermions in Table~\ref{brsir}.

\subsection{Dynamical Higgs phase in the $\{\A,N,p\}$ models     \label{brokenA} }

In  the $\{\A,N,p\}$ model there is  a competition between two possible bifermion symmetry breaking channels $\chi \eta$ and $  \xi \eta $. 
This time,   the MAC criterion would favor the  $  \xi \eta $  condensates against  $\chi \eta$. Indeed,      
the strength of the one-gluon exchange potential for the two channels
 \bea   &     \chi \left({\raisebox{-9pt}{\yng(1,1)}}\right) \,  \,  \eta \Big(\bar{\raisebox{-2pt}{\yng(1)}}\Big)      \qquad  \   & {\rm forming}    \quad      \raisebox{-2pt}{\yng(1)}\;,\nonumber    \\
&     \xi \Big({\raisebox{-2pt}{\yng(1)}}\Big) \, \eta \Big(\bar{\raisebox{-2pt}{\yng(1)}}\Big)    \qquad \ & {\rm forming}  \quad  (\cdot)  \;,
 \eea
are, respectively,
 \bea       \frac{N^2-1}{2N}  -      \frac{ (N-2)(N+1)}{N}  -   \frac{N^2-1}{2N}  &=&  -      \frac{ (N-2)(N+1)}{N} \;,\nonumber \\
     0 -   2       \frac{N^2-1}{2N}   & =&  -       \frac{N^2-1}{N}  \;.
 \eea
Again, these perturbative estimates are not excessively significant, and  we assume that
 both  condensates occur as:
\bea
&& \brc  \chi^{ij}   \eta_i^A \ckt  =\,   c_{\chi\eta} \,  \Lambda^3   \delta^{j A}\ne 0\;,   \qquad   j=1,\dots,  N-4\;,   \quad  A=1, \dots , N -4 \;,  \nonumber \\
&&  \brc  \xi^{i,a}   \eta_i^B \ckt =\,   c_{\eta\xi} \,  \Lambda^3   \delta^{aB}\ne 0\;,   \qquad  a=1,\dots,  p \;,   \quad B=N-4+1,\dots,N-4+  p \;. 
  \eea
The pattern of symmetry breaking is
\bea
&& SU(N)_{\rmc}  \times   SU(N-4+p)_{\eta}  \times  SU(p)_{\xi}  \times  U(1)_{\chi\eta}\times  U(1)_{\chi\xi} \nn \\
&&  \xrightarrow{\brc  \xi   \eta \ckt , \brc  \chi \eta \ckt}      SU(4)_{{\rm  c}}  \times  SU(N-4)_{{\rm cf}_{\eta}}  \times   SU(p)_{\eta\xi}  \times  U(1)_{\chi \eta}^{\prime} \times  U(1)_{\chi \xi}^{\prime} \;.
\label{symbrea}
\eea

The color gauge symmetry is partially (dynamically) broken, leaving  color-flavor diagonal  global $SU(N-4)_{{\rm cf}_{\eta}} $ symmetry and an $SU(4)_{{\rm  c}}$ gauge symmetry.
$U(1)_{\chi \eta}^{\prime}$ and $U(1)_{\chi \xi}^{\prime}$ are a combinations respectively  of $U(1)_{\chi \eta}$ (\ref{uce}) and $U(1)_{\chi \xi}$ (\ref{ucx}) with  the elements of $SU(N)_{\rmc} $ and $ SU(N-4+p)_{\eta}$ generated by:
\bea  
&& \qquad  t_{SU(N)_{\rm c}}=  \left(\begin{array}{c|c}
2 \frac{  \alpha  (N-4+p ) +\beta p }{N-4}  {\bf 1}_{(N-4)\times (N-4)}&\\
\hline
& - \frac{  \alpha  (N-4+p ) +\beta p }{2}  {\bf 1}_{4\times 4}\\
\end{array}\right) \ , \nn \\
&& t_{SU(N-4+p)_{\eta}}=  \left(\begin{array}{c|c}
-\frac{p (\alpha + \beta ) (N-2 )}{N-4} {\bf 1}_{(N-4)\times (N-4)}&\\
\hline
&  (\alpha + \beta ) (N-2 )  {\bf 1}_{p\times p}\\
\end{array}\right) \;.  
\eea
Making the decomposition of the fields in the direct sum of representations in  the subgroup one arrives at  Table~\ref{brauv}.
\begin{table}[h!t]
{  \centering 
  \small{\begin{tabular}{|c|c |c|c|c|c|  }
\hline
\su       &    $SU(N-4)_{{\rm cf}_{\eta}}   $  &  $SU(4)_{{\rm c}}$ &  $ SU(p)_{\eta\xi}$ &   $  U(1)_{\chi \eta}^{\prime}$   &   $ U(1)_{\chi \xi}^{\prime}$ \\
\hline
\sbbuu $\chi_1$     &       $ { \yng(1,1)} $ & $  \frac{(N-4)(N-5)}{2} \cdot   (\cdot) $  & $ \frac{(N-4)(N-5)}{2} \cdot   (\cdot)   $   &  $\frac{(N-4+p) N}{(N-4)}    $  &  $p \frac{N}{N-4}$  \\
$\chi_2$                &   $ 4 \cdot { \yng(1)} $  &    $  (N-4)  \cdot    { \yng(1)} $    & $ 4(N-4) \cdot   (\cdot)   $   &$\frac{(N-4+p) N}{2(N-4)}  $  &  $\frac{p N}{2(N-4)}$  \\
$\chi_3$                &    $  6  \cdot   (\cdot) $   &   $ { \yng(1,1)} $   & $ 6  \cdot   (\cdot)   $   &  $0$  & $0$  \\
$ \eta_1$                 &  $  {\bar  {\yng(2)}} \oplus {\bar  {\yng(1,1)}}  $     &   $(N-4)^2  \cdot   (\cdot )    $   &   $ (N-4)^2  \cdot  (\cdot ) $   & $ -\frac{(N-4+p) N}{(N-4)} $  & $-\frac{p N}{N-4} $\\   
$ \eta_2$             & $ p  \cdot \bar{\yng(1)} $      &   $p(N-4)  \cdot  (\cdot )  $      &   $ (N-4)  \cdot \bar{\yng(1)} $    &$-2 -2 \frac{p}{N-4}$ & $N-2 -\frac{2 p}{N-4}  $\\
$ \eta_3$              &   $  4   \cdot   {\bar  {\yng(1)}}$     & $ (N-4)  \cdot   {\bar  {\yng(1)}} $     &   $ 4  (N-4)     \cdot   (\cdot )  $    &$-\frac{(N-4+p) N}{2(N-4)} $ &$ -\frac{p N}{2(N-4)} $ \\
$ \eta_4$              &   $ 4 p \cdot  (\cdot ) $     & $ p \cdot   {\bar  {\yng(1)}}   $     &   $4  \cdot \bar{\yng(1)}   $    &$\frac{N-4+p }{2}$ &$N-2  + \frac{p }{2}$ \\
$ \xi_1$                &   $  p  \cdot   { {\yng(1)}}    $     & $p(N-4)   \cdot  (\cdot )  $     &   $ (N-4)   \cdot   { {\yng(1)}}   $   & $2 + 2\frac{p}{N-4}$  & $-(N-2)+\frac{2p}{N-4}$\\
$ \xi_2$                &   $4 p   \cdot  (\cdot )   $     & $  p  \cdot   { {\yng(1)}}    $     &   $ 4  \cdot   { {\yng(1)}}   $   & $-\frac{N-4+p}{2}$  & $-(N-2)-\frac{p}{2}$\\
\hline 
\end{tabular}}  
  \caption{\footnotesize   UV fieds in the $\{\A,N,p\}$ model, decomposed as a direct sum of the representations of the unbroken group of  Eq.~(\ref{symbrea}). }
\label{brauv}
}
\end{table}

\begin{table}[h!t]
{  \centering 
  \small{\begin{tabular}{|c|c |c|c|c|  }
\hline
\su     &    $SU(N-4)_{{\rm cf}_{\eta}}   $    &  $ SU(p)_{\eta\xi}$ &   $  U(1)_{\chi \eta}^{\prime}$   &   $ U(1)_{\chi \xi}^{\prime}$ \\
 \hline
\sbbu $ {\cal B}   $          &  $  {\bar  {\yng(2)}}  $     &   $\frac{(N-4)(N-3)}{2}\  \cdot  (\cdot ) $   &  $ -\frac{(N-4+p) N}{(N-4)} $  & $-\frac{p N}{N-4} $\\
  \hline 
\end{tabular}}  
  \caption{\footnotesize   IR fied in the $\{\A,N,p\}$ model in the dynamical Higgs phase. }
\label{brair}
}
\end{table}
The composite massless baryons are subset of those in (\ref{baryons20}): 
\bea
 && {\cal B}^{\{AB\}} =  \chi^{ij}   \eta_{i}^{A}  \eta_{j}^{B} \;,  \qquad  A,B = 1, \dots, N-4 \;.   \label{baryonbra}
\eea
In the IR these fermions  saturate all the anomalies of the unbroken chiral symmetry. This can be seen by  an inspection of  Table~\ref{brair} and   Table~\ref{brauv},    with the help of the following observation.   

In fact,   there is a novel feature in the  $\{\A,N,p\}$ models, which is not shared by the  $\{\S,N,p\}$ models.   As  seen in  Table~\ref{brair},
there is an  unbroken strong gauge symmetry $SU(4)_{{\rm c}}$, with a  set of  fermions, 
\be  \chi_3\;,\quad \chi_2\;,\quad  \eta_3\;,\quad \eta_4\;,\quad \xi_2\;,\label{these}
\ee
charged with respect to it.     However,   the pairs  $\{ \chi_2\;,\,  \eta_3 \}$  and   $\{ \eta_4\;,\,  \xi_2 \}$ can form massive Dirac fermions and decouple.  These are vectorlike  with respect to the  surviving infrared symmetry, (\ref{symbrea}), hence are irrelevant to the anomalies.\footnote{Actually,   
  with matter fermions (\ref{these})  $SU(4)_{{\rm c}}$  is asymptotically free only for    $50- 2N - 2 p  > 0$.   If     $50- 2N - 2 p  <  0$,    $SU(4)_{{\rm c}}$  will remain weakly coupled in the infrared, 
but the fact that the fermions (\ref{these}) do not contribute to the anomalies with respect to  the remaining flavor symmetries  (\ref{remainingfl})  stays valid.}
   On the other hand,   the fermion $\chi_3$  can condense 
\be    \brc  \chi_3 \chi_3 \ckt  \
\ee
forming massive composite mesons, $\sim \chi_3 \chi_3$, which also decouples.  It is again neutral with respect to all of 
\be SU(N-4)_{{\rm cf}_{\eta}}  \times   SU(p)_{\eta\xi}  \times  U(1)_{\chi \eta}^{\prime} \times  U(1)_{\chi \xi}^{\prime} \;.\label{remainingfl}
\ee
To summarize,    $SU(4)_c$  is invisible  (confines) in the IR, and  only the unpaired  part of the $\eta_1$  fermion   $\big({\bar  {\yng(2)}}\big)$  remains  massless, and its contribution to the anomalies is reproduced exactly by the composite fermions, (\ref{baryonbra}).  

~~~~~

{\noindent  \bf Comment:}
The massive mesons  $\{ \chi_2\,  \eta_3 \}$,    $\{ \eta_4 \,  \xi_2 \}$,  $ \{\chi_3 \, \chi_3\}$  are not charged with respect to the flavor symmetries surviving in the infrared.  
It is tempting to regard them as a toy-model  ``dark matter",  as contrasted to  the fermions  $ {\cal B}^{AB} $ which constitute the ``ordinary, visible" sector.


\section{Symmetries \label{symmetries} }

In the conventional 't Hooft anomaly analysis discussed above   only the algebra of the group matters.  
In this section the symmetry of the models will be  examined with more care, by taking into account the global aspects of the color and flavor symmetry groups. 
Let us first consider  the  Bars-Yankielowicz ($\{\S,N,p\}$) models.

For a $\{\S,N,p\}$ model,   the classical symmetry group of our system is given by 
\bea   G_{\mathrm{class}}
&=& G_{\mathrm{c}} \times G_{\mathrm{f}} \nonumber \\ 
&= &SU(N)_{\mathrm{c}} \times  \frac{       U(1)_{\psi} \times  U(N+4+p)_{\eta} \times  U(p)_{\xi}  }{ \mathbb{Z}_{N}} \;. 
\label{eq:symmetry_classical_psieta}
\eea
The color group is $G_{\mathrm{c}} = SU(N)_{c} $,  and its center acts non-trivially on the matter  fields: 
\be   
 \mathbb{Z}_{N}:  \psi \to  \rme^{  \frac{4\pi \im n }{N}} \psi \;, \qquad     \eta  \to  \rme^{-\frac{ 2 \pi \im n}{N}} \eta  \;, \qquad     \xi  \to  \rme^{\frac{ 2 \pi \im n}{N}} \xi  \;, \label{zn}
\ee
($n \in {\mathbbm Z}$).
The division by $\mathbb{Z}_N$ in Eq.~(\ref{eq:symmetry_classical_psieta})  is due to the fact that the numerator  overlaps with the center of the gauge group (see 
Sec.~\ref{ZNinH} below).
Another, equivalent way of writing the flavor part of the classical symmetry group is
\be   G_{\mathrm{f}}=  \frac{ SU(N+4+p) \times  SU(p)  \times U(1)_{\psi} \times U(1)_{\eta}   \times U(1)_{\xi} }{ \mathbb{Z}_{N}  \times \mathbb{Z}_{N+4+p} \times \mathbb{Z}_{p}} \;. 
\label{scv}
\ee

Quantum mechanically one must consider the effects of the anomalies and $SU(N)$ instantons which reduce the flavor group down to its anomaly-free subgroup. 
The instanton vertex explicitly breaks the three  independent $U(1)$ rotations for $\psi$, $\eta$ and $\xi$ down to two $U(1)$'s,  to be chosen among
$U(1)_{\psi\eta}$, $U(1)_{\psi\xi}$, and  $U(1)_{\xi\eta}$:
    \bea
U(1)_{\psi\eta} &:& \qquad   \psi\to \rme^{\im (N+4+p)\alpha}\psi\;, \quad  \eta \to \rme^{-\im (N+2)\alpha}\eta\;, \nonumber \\
U(1)_{\psi\xi} &:& \qquad  \psi\to \rme^{\im p \beta}\psi\;, \quad  \xi \to \rme^{-\im (N+2)\beta}\xi\;, \nonumber \\
U(1)_{\eta\xi} &:& \qquad  \eta \to \rme^{\im p \gamma}\eta \;, \quad  \xi \to \rme^{-\im (N+4+p)\gamma}\xi\;
\label{upexBis}
\eea  
(see Eq.~(\ref{upe})-Eq.~(\ref{upex})). Three different discrete sub-groups left unbroken are 
\beq
\Zp :  \  \psi \to \rme^{\frac{2\pi \im k }{ N+2}} \psi\;,  \qquad
\Zes : \    \eta\to \rme^{\frac{2\pi \im k }{ N+4+p}}\eta\;,   \; \qquad  
\Zx : \    \xi\to \rme^{\frac{2\pi \im k }{ p}}\xi\;.
\label{dis}
\eeq
The question is:  which is the correct anomaly-free sub-group? The anomaly affects only the $U(1)$ part of the group
\be
U(1)_{\psi} \times U(1)_{\eta}   \times U(1)_{\xi}  \xrightarrow{\rm anomaly}   {\cal H} 
\ee
so that the total symmetry group is broken as follows
\be   G_{\mathrm{f}} \xrightarrow{\rm anomaly}\frac{  SU(N+4+p) \times  SU(p) \times {\cal H}  }{ \mathbb{Z}_{N}  \times \mathbb{Z}_{N+4+p} \times \mathbb{Z}_{p}} \;. 
\label{scv}
\ee

\subsection{Study of $ {\cal H} $}

 Clearly, $U(1)_{\psi\eta}$,  $U(1)_{\psi\xi}$, $ U(1)_{\eta\xi}$, $  \Zp$, $\Zes$, $\Zx$  are all  part of the 
 anomaly-free sub-group, but one must find the minimal description,  in order to avoid the double-counting. 
${\cal H}$ is at the bottom of  the following sequence of covering spaces:
\beq
\begin{array}{c}
 U(1)_{\psi\eta} \times  U(1)_{\psi\xi}   \times  U(1)_{\eta\xi}  \times \Zp \times
\Zes \times
\Zx \\
\downarrow
\\
 U(1)_{\psi\eta} \times  U(1)_{\psi\xi}   \times  \Zp \\
\downarrow \\
 {\cal H}
\end{array}
\eeq
The first arrow can be understood as follows. $U(1)_{\eta\xi}$ can always be obtained by a combination of the other two continuous groups, by choosing (using conventions for $\alpha$, $\beta$, $\gamma$ as in Eq.~(\ref{upexBis}))
\beq
\alpha = - \frac{p \gamma}{N+2} \;, \qquad  \beta = \frac{(N+4+p) \gamma}{N+2} \ .
\eeq
Also, the fundamental element of $\Zes$ can  be obtained by a combination of the fundamental of   $\Zp$ ($k=1$ in Eq.~(\ref{dis})) with the   $U(1)_{\psi\eta}$ 
element 
\beq
\alpha = - \frac{1}{(N+4+p)(N+2)} \ .
\eeq
Similarly $\Zx$ can always be expressed as part of  $U(1)_{\psi\xi} \times \Zp$.

The question now (the second arrow)  is  whether 
 \be     ({\mathbbm Z}_{N+2})_{\psi}   \subset  U(1)_{\psi\eta} \times  U(1)_{\psi\xi}         \label{ZN+2}
\ee
holds, i.e.,   whether  the discrete part of the group can be entirely expressed as a  subgroup  of the continuous $U(1)$ groups.  
The requirement (\ref{ZN+2}) is equivalent to
\bea    (N+ 4 +p)   { \alpha}+ p \, {  \beta} &\sim & \frac{2\pi}{N+2} \;, \nonumber \\
   - (N+2)  {  \alpha} &\sim& 0\;,  \nonumber \\
   - (N+2)   { \beta}  & \sim & 0  \;, \label{tosolvebis!!}
\eea
where $\sim$ means the equality with possible additional terms  of the form $2\pi \times $\,  integer allowed. 
It follows from the last  two equations that 
\be   { \alpha}  =  \frac{2\pi m }{N+2} \;,  \qquad    {  \beta}  =  \frac{2\pi n }{N+2} \;, \qquad  m,n \in {\mathbbm Z}\;,
\ee 
which inserted in the first gives
\be   \frac{2\pi m    (N+ 4 +p)   }{N+2}+    \frac{2\pi n   p    }{N+2}   \sim   \frac{2\pi}{N+2} \;,
\ee
that is,
\be    (2 +p) m + n   p    =   1 +       (N+2) \ell\;,\qquad  m,n,\ell \in {\mathbbm Z}\;.\label{this}
\ee
If one (or both) of $N$ and $p$ is odd, Eq.~(\ref{this})  has solutions.  That is Eq.~(\ref{ZN+2}) is valid, 
and  $ {\cal H}$ has only one component connected to the identity. This also means that, in the context of the conventional anomaly matching discussion,  
 the anomaly matching requirement involving $({\mathbbm Z}_{N+2})_{\psi}$,  $\Zes$, or  $\Zx$  is automatically satisfied  when the  triangles containing  $U(1)_{\psi\eta} \times  U(1)_{\psi\xi}    \times  SU(N+p+4)\times SU(p) $  are UV-IR matched.
 
Vice versa, if $p$ and $N$ are  both   even there are no solutions of Eq.~(\ref{this}):   i.e., 
 $ ({\mathbbm Z}_{N+2})_{\psi}  $ is not entirely contained in  $ U(1)_{\psi\eta} \times  U(1)_{\psi\xi}    $;  only the even elements of   $({\mathbbm Z}_{N+2})_{\psi}$  are:
\be     \big({\mathbbm Z}_{\frac{N+2}{2}}\big)_{\psi}    \subset  U(1)_{\psi\eta} \times  U(1)_{\psi\xi}    \;.     \label{ZN+2again}
\ee
One can show however that for $p$, $N$ both   even 
 \be     ({\mathbbm Z}_{N+2})_{\psi}  \subset  U(1)_{\psi\eta} \times  U(1)_{\psi\xi}    \times  ({\mathbbm Z}_2)_F \;,     \label{ZN+2Z2}
\ee
where $\Zf$  is the fermion parity generated by
\beq
\psi \to -\psi \;, \qquad \eta \to -\eta \;, \qquad \xi \to -\xi \;.\label{fp}
\eeq
In fact, admitting the presence of fermion parity the requirement   (\ref{tosolvebis!!}) gets modified to 
\bea    (N+ 4 +p)   { \alpha}+ p \, {  \beta} &\sim & \frac{2\pi}{N+2} +\pi  \;, \nonumber \\
   - (N+2)  {  \alpha} &\sim& \pi \;,  \nonumber \\
   - (N+2)   { \beta}  & \sim & \pi    \;,
\eea
and thus 
\be    (2 +p) m + n   p    =       (N+2) \ell\;,\qquad  m,n,\ell \in {\mathbbm Z}\;.\label{thisBis}
\ee
which always has a solution.

To summarize,  when $p$ and  $N$ are both   even,   one has
\be   { \cal H } =   U(1)_1 \times  U(1)_2   \times  ({\mathbbm Z}_2)_F \;,
\ee
i.e. it has two disconnected components.  $U(1)_1$ and  $U(1)_2$  are any two out of 
$U(1)_{\psi\eta}$,  $U(1)_{\psi\xi}$, and   $U(1)_{\eta\xi}$.  If  $p$ and/or $N$ is odd, instead,  
\be  {  \cal H  } =   U(1)_1 \times  U(1)_2\;:
\ee
it has only one connected component.

\subsection{ ${\mathbbm Z}_N   \subset {\cal H}$      \label{ZNinH}}

We focus now on the center of   the color  $SU(N)$ group,  
${\mathbbm Z}_N $. 
We first  show  that when $N$, $p$ are both even,  
\be    {\mathbbm Z}_N \not \subset  U(1)_{\psi\eta} \times  U(1)_{\psi\xi}   \;.  
\ee
To prove this, {\it  ab absurdo},   assume that $U(1)_{\psi\eta} \times  U(1)_{\psi\xi}  $   does contains  $  {\mathbbm Z}_N$: that is
\bea 
  (N+ 4 +p)   {  \alpha}+ p \, {  \beta} &\sim&  \frac{4\pi}{N}  \;,\nonumber \\
  - (N+2)  {  \alpha} &\sim&   -   \frac{2\pi}{N} \;,\nonumber \\
   - (N+2)   { \beta}     &\sim&   \frac{2\pi}{N}  \;. 
\label{notsolve}
\eea  
(Remember that the symbol ``$\sim$"   here indicates equality modulo terms of the form $2\pi n$, $n \in {\mathbbm Z}$.)
We first eliminate    ${  \alpha}$  from the first two.   As $N$, $p$  are both even,   multiply the first by  $\tfrac{N+2}{2}$  and the second by $\tfrac{N+4 +p}{2}$
(both integers) 
and add.   We get
\be     \frac{p}{2}       (N+2) { \beta}   \sim   \frac{4\pi}{N}    \frac{N+2}{2} -    \frac{2\pi}{N}   \frac{N+4 +p}{2} 
 \sim     \pi  -   \frac{ \pi p}{ N}\;.
\label{contradction1}
\ee
On the other hand   multiplying the third of Eq.~(\ref{notsolve}) by  $\frac{p}{2}$ (also an integer)  gives 
\be  \frac{p}{2}    (N+2) { \beta}     \sim -   \frac{ \pi p}{ N}\;.\label{contradction2}
\ee
Eq.~(\ref{contradction1}) and Eq.~(\ref{contradction2})  contradict each other.       
 {\it Q.E.D.}

We next prove that if at least one of      $N$  and    $p$ is odd,   then
\be    {\mathbbm Z}_N     \subset  U(1)_{\psi\eta} \times  U(1)_{\psi\xi}   \;, 
\ee
that is,   Eq.~(\ref{notsolve}) has solutions.  
To  prove this,    we repeat the procedure above,   noting that there may be  now
extra terms on the right hand side.   As a result,  Eq.~(\ref{contradction1})  is replaced by
\be     \frac{p}{2}    (N+2) {  \beta}     =         \pi  -   \frac{2\pi p}{2N}  +   2\pi m  \frac{N+2}{2} + 2\pi n  \frac{N+4+p}{2} \;,\label{contra1rep} 
\ee
while  Eq.~(\ref{contradction2})  is replaced by
\be  \frac{p}{2}    (N+2) {\tilde \beta}    =    -   \frac{2\pi p}{2N} +  2 \pi \ell \cdot \frac{p}{2}\;, \label{contra2rep}
\ee
\be   m, n, \ell   \in  {\mathbbm Z}\;.
\ee
Now when one or both of   $N$  and    $p$  is odd, it is always possible to find appropriate  integers  $m, n, \ell $   such that  
the right hand sides of Eq.~(\ref{contra1rep}) and  Eq.~(\ref{contra2rep})  are equal, that is, 
\be      \pi  +   2\pi m  \frac{N+2}{2} + 2\pi n  \frac{N+4+p}{2}   \sim    2 \pi \ell \cdot \frac{p}{2}\;.   \label{clearly}
\ee
When  both $N$ and $p$ are even,  exceptionally,  this  equality  does not hold for any choice of $m, n, \ell $, as has been already noted.

Finally,   we  prove that
\be    {\mathbbm Z}_N     \subset  U(1)_{\psi\eta} \times  U(1)_{\psi\xi}  \times   ({\mathbbm Z}_2)_F \;, 
\ee
when  $N$ and $p$ are both even.  This means that     (cfr.   Eq.~(\ref{notsolve}))
\bea 
  (N+ 4 +p)   {  \alpha}+ p \, {  \beta} &\sim&  \frac{4\pi}{N}  + \pi  \;,\nonumber \\
  - (N+2)  {  \alpha} &\sim&   -   \frac{2\pi}{N}  + \pi   \;,\nonumber \\
   - (N+2)   { \beta}     &\sim&   \frac{2\pi}{N}   +    \pi  \;.
\label{notsolveMod}
\eea  
 Let us repeat the procedure  Eq.~(\ref{notsolve})-Eq.~(\ref{contradction2}),  by keeping the extra terms coming from  $ \pi$ on the right hand sides.  
Eq.~(\ref{contradction1})  is replaced by
\be      \frac{p}{2}       (N+2) { \beta}   \sim   \pi  -   \frac{ \pi p}{ N}  +  \frac{p+2}{2} \pi\;, \label{nocontra1}
\ee
whereas  Eq.~(\ref{contradction2})  is modified to   
\be  \frac{p}{2}    (N+2) { \beta}     \sim -   \frac{ \pi p}{ N}    +   \frac{p}{2} \pi  \;.\label{nocontra2}
\ee
The right hand sides of  Eq.~(\ref{nocontra1}) and  Eq.~(\ref{nocontra2})  now agree.

To sum up, we  have  shown  that 
\be   {\mathbbm Z}_N   \subset {\cal H} 
\ee
for any choice of  $N$ and $p$,   for  the  $\{\S,N,p\}$ models.

\subsection{ $\{\A,N,p\}$ models}

So far,  our analysis concentrated on the   $\{\S,N,p\}$ models  for definiteness.
For the $\{\A,N,p\}$ models,   the result is very similar. The symmetry group is
\be   G_{\mathrm{f}}=  \frac{ SU(N-4+p) \times  SU(p) \times U(1)_{\chi} \times U(1)_{\eta}   \times U(1)_{\xi}  }{ \mathbb{Z}_{N}  \times \mathbb{Z}_{N-4+p} \times \mathbb{Z}_{p}} \;,
\ee
where the anomaly acts on the $U(1)$ part as
\be
U(1)_{\chi} \times U(1)_{\eta}   \times U(1)_{\xi}  \xrightarrow{\rm anomaly}   {\cal H}\;. 
\ee
Clearly all $U(1)_{\chi\eta}$,  $U(1)_{\chi\xi}$, $ U(1)_{\eta\xi}$ defined in Eq.~(\ref{uce})-Eq.~(\ref{ucex}) together with the discrete groups 
\beq
\Zc :  \  \chi \to \rme^{\frac{2\pi \im k }{ N-2}} \chi\;,  \qquad
\Zea : \    \eta\to \rme^{\frac{2\pi \im k }{ N-4+p}}\eta\;,   \; \qquad  
\Zx : \    \xi\to \rme^{\frac{2\pi \im k }{ p}}\xi\;.
\eeq
are the nonanomalous symmetry group of the system,  but  we need a minimum set without redundancy. 
For  $p=0$, the $\chi\eta$ model, the result is:
\bea  && N \  {\rm odd}: \qquad   {\cal H}=  U(1)_{\chi\eta} \;, \nn \\
   && N \ {\rm even}: \qquad   {\cal H}=  U(1)_{\chi\eta} \times \Zf   \;.
\eea
For greater $p$, as for the $\{\S,N,p\}$ model,  $ {\cal H}$   is:
 \bea  
&&{\rm gcd}(N,p,2)=1 : \qquad {\cal H} =  U(1)_{1}   \times  U(1)_{2}   \nn \;, \\ 
 && {\rm gcd}(N,p,2)=2:  \qquad {\cal H} =  U(1)_{1}   \times  U(1)_{2}  \times  ({\mathbbm Z}_2)_F  \;.  
 \eea
 where   $U(1)_{1,2}$ are any two out of  $U(1)_{\chi\eta}$,  $U(1)_{\chi\xi}$ and  $ U(1)_{\eta\xi}$.
Again,
\be   {\mathbbm Z}_N   \subset {\cal H} 
\ee
for any choice of  $N$ and $p$.  The proof for the  $\{\A,N,p\}$ models  is entirely analogous  to the one  given for  $\{\S,N,p\}$  and is omitted.

\subsection{Illustration}

Let us illustrate the symmetry of our systems graphically,  taking a few concrete models of 
 the  type,  $\{\S,N,p\}$.

It is convenient to introduce the following notation. We parameterize a generic 
$U(1) \subset  T^3 = U(1)_{\psi} \times U(1)_{\eta}   \times U(1)_{\xi}$ with a triplet of integer numbers
\be{\bf t}  = \left(\begin{array}{c}t_1\\t_2\\t_3\end{array}\right) \in \Z^3\;, \ee
so that
\be
 U(1): \ \left(\begin{array}{c}\psi\\\eta\\\xi \end{array}\right) \rightarrow  \left(\begin{array}{c}e^{i t_1 \theta } \psi\\e^{i t_2 \theta } \eta\\e^{i t_3 \theta } \xi \end{array}\right) \;, \qquad 0\leq \theta <2\pi\;.
\ee
This $U(1)$ winds ${\rm gcd}(t_1,t_2,t_3)$-times around the three-torus $T^3$. In general, given a specific direction, we choose the ``fundamental'' generator for which ${\rm gcd}(t_1,t_2,t_3) = 1$ so that periodicity in $\theta$ is exactly $2\pi$. In this notations the three fundamental $U(1)$'s are generated by
\be
{\bf t}_{ \scriptscriptstyle U(1)_{\psi} } = \left(\begin{array}{c}1\\0\\0\end{array}\right)\;, \qquad {\bf t}_{\scriptscriptstyle U(1)_{\eta}}= \left(\begin{array}{c}0\\1\\0\end{array}\right)\;, \qquad {\bf t}_{ \scriptscriptstyle U(1)_{\xi}} \left(\begin{array}{c}0\\0\\1\end{array}\right) \;,
\ee
and the non-anomalus ones are generated by
\be
{\bf t}_{\scriptscriptstyle U(1)_{\psi\eta} } = \left(\begin{array}{c}\frac{N+4+p}{{\rm gcd} (N+4+p,N+2)}\\-\frac{N+2}{{\rm gcd}(N+4+p,N+2)}\\0\end{array}\right), \  {\bf t}_{\scriptscriptstyle U(1)_{\psi\xi}}= \left(\begin{array}{c}\frac{p}{{\rm gcd} (p,N+2)}\\0\\-\frac{N+2}{{\rm gcd} (p,N+2)}\end{array}\right), \  {\bf t}_{\scriptscriptstyle U(1)_{\eta\xi}} = \left(\begin{array}{c}0\\\frac{p}{{\rm gcd}(N+4+p ,p)}\\-\frac{N+4+p}{{\rm gcd} (N+4+p ,p)}\end{array}\right)\;.    \label{reallycorrect} 
\ee
We give now specific examples for $p=0,1,2$. 
\begin{itemize}
\item
For  {$p=0$},  the $\psi\eta$ model,  this  has been discussed in detail in \cite{BKL2} and the result is:
\bea  && N \ \ {\rm odd}: \qquad {\cal H}=  U(1)_{\psi\eta}\;,    \nn \\
   && N \ {\rm even}: \qquad  {\cal H} =   U(1)_{\psi\eta} \times \Zf \;.   
\eea

\item
For  {$p=1$},  independentely on $N$, $ {\cal H} $ has only one connected component. In Figure \ref{S31} we show the case $N=3$.
One possible way to parameterize   ${\cal H}$  is
\bea  &&  {\cal H} =   U(1)_{\psi\xi}   \times  U(1)_{\eta\xi}\;.   
\eea
Note that $U(1)_{\psi\xi} $ contains $({\mathbbm Z}_{N+2})_{\psi}$ and $U(1)_{\eta\xi}$ contains   $({\mathbbm Z}_{N+5})_{\eta}$, so together they contain the whole discrete lattice  $({\mathbbm Z}_{N+2})_{\psi} \times ({\mathbbm Z}_{N+5})_{\eta}$.
We can define the group $\tU(1)$ as the one that contains $\Z_N$, and is the one generated by
\be
{\bf t}_{\scriptscriptstyle \tU(1)} = 2 {\bf t}_{\scriptscriptstyle U(1)_{\psi\xi}} - {\bf t}_{\scriptscriptstyle U(1)_{\eta\xi} } 
\label{tu1}
\;.
\ee
\begin{figure}[h!]
\centering
\includegraphics[width=1.05\linewidth]{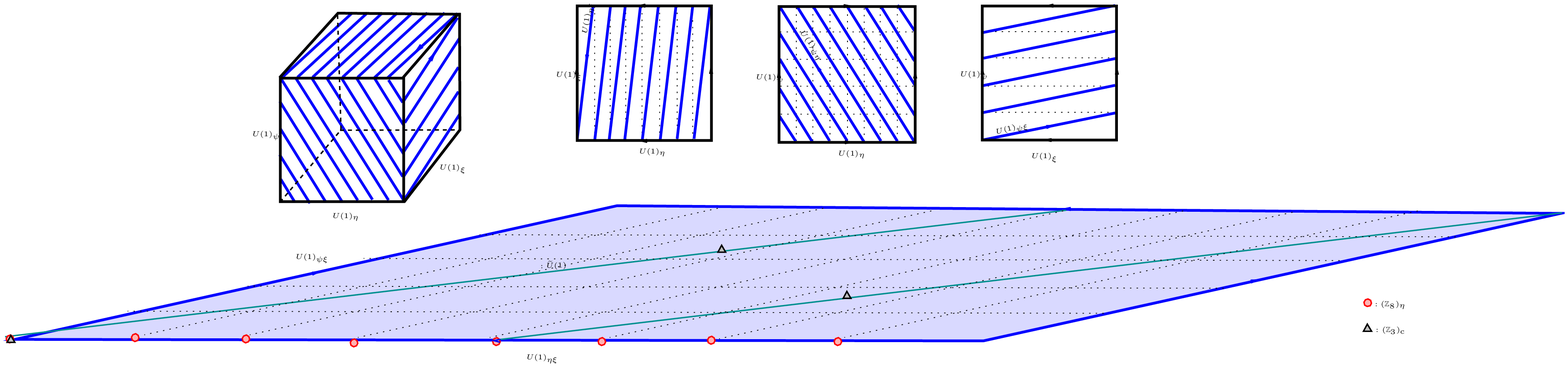}
\caption{\footnotesize The three-torus $U(1)_{\psi} \times U(1)_{\eta} \times U(1)_{\xi}$ broken to $U(1)_{\psi\xi}   \times  U(1)_{\eta\xi}$ for the $\{\S,3,1\}$ model. }
\label{S31}
\end{figure}

\item
For  {{$p=2$, $N$ odd}},  $ {\cal H} $ has only one connected component. In Figure \ref{S32} we show the graphs for the case $N=3$.
One possible way to parameterize   {\cal H} is
\bea  &&  {\cal H} =   \frac{U(1)_{\psi\xi}   \times  U(1)_{\eta\xi}}{\Z_2}   
\label{hz2}
\eea
Note that $U(1)_{\psi\xi} $ contains $({\mathbbm Z}_{N+2})_{\psi} \times ({\mathbbm Z}_{2})_{\xi} $ 
and $U(1)_{\eta\xi}$ contains   $({\mathbbm Z}_{N+6})_{\eta} \times ({\mathbbm Z}_{2})_{\xi} $ 
 so    $U(1)_{\psi\xi} \times U(1)_{\eta\xi} $ contains ${\mathbbm Z}_{2(N+2)} \times {\mathbbm Z}_{2(N+6)}$ which  twice redundant with respect to $({\mathbbm Z}_{N+2})_{\psi} \times ({\mathbbm Z}_{N+6})_{\eta} \times ({\mathbbm Z}_{2})_{\xi} $. 
 We can also see this in the following way. $U(1)_{\psi\xi}$ contains a non-trivial element of $U(1)_{\eta\xi}$. If we take the element of  $U(1)_{\psi\xi}$ with $\beta= \pi$ we obtain
\beq
\psi \to \psi \ , \quad \eta \to \eta \ , \quad \xi \to - \xi\ 
\eeq
which is exactly the element of $U(1)_{\eta\xi}$ with $\gamma = \pi$. This is the reason for  the $\Z_2$ division in (\ref{hz2}).
The group $\tU(1)$   that contains $\Z_N$  is the one generated by (\ref{tu1}).
If we define $ \hU(1)$  generated by
\beq
 {\bf t}_{\scriptscriptstyle \hU(1)}   =  -\frac12 {\bf t}_{\scriptscriptstyle U(1)_{\psi\xi}} +    \frac12   {\bf t}_{\scriptscriptstyle U(1)_{\eta\xi} } \;,
\eeq
we can write
\beq
 {\cal H} =    U(1)_{\psi\xi}  \times   \hU(1)  =  U(1)_{\eta\xi}  \times   \hU(1) \;.
\eeq
\begin{figure}[h!]
\centering
\includegraphics[width=1.05\linewidth]{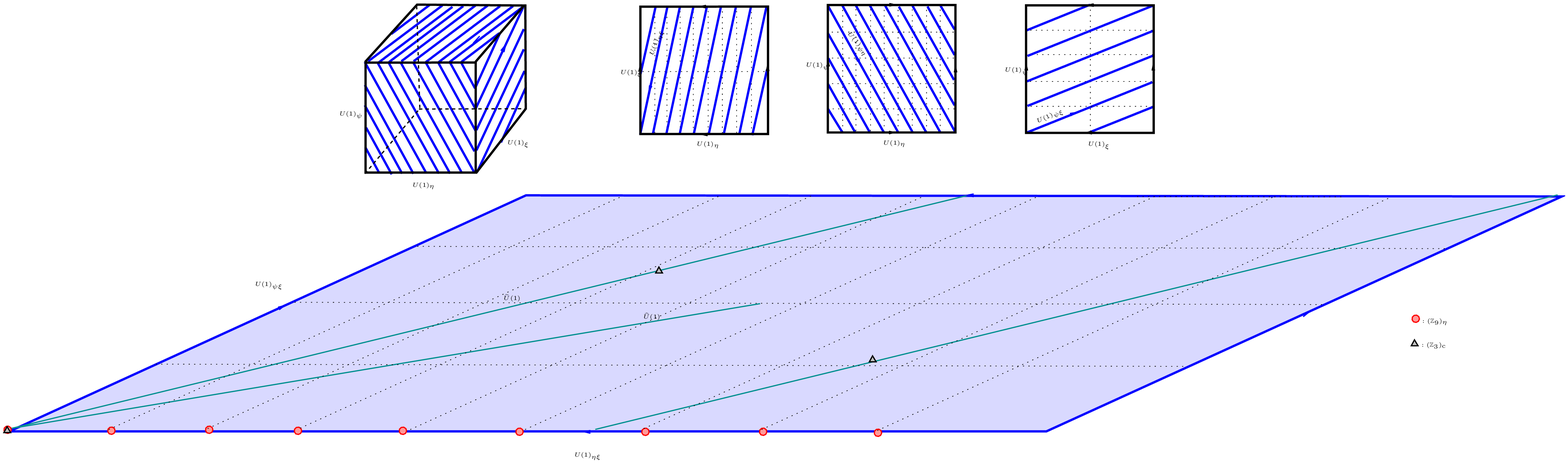}
\caption{\footnotesize The three-torus $U(1)_{\psi} \times U(1)_{\eta} \times U(1)_{\xi}$ broken to $\frac{U(1)_{\psi\xi}   \times  U(1)_{\eta\xi}}{\Z_2}$  for the $\{\S,3,2\}$ model. }
\label{S32}
\end{figure}

\item
For  {{$p=2$, $N$ even}}, $ {\cal H} $ has two components. In Figure \ref{S42},  we illustrate  the case $N=4$, $p=2$.
One possible way to parameterize   ${\cal H}$  is
\bea  &&  {\cal H} =  U(1)_{\psi\xi}   \times  U(1)_{\eta\xi}   \times \Zf  \;.
\eea
We can define the group $\tU(1)$  generated by (\ref{tu1}) but this time it contains only  $ \Z_{\frac{N}{2}}$. In general it is not possible to write $\Z_{{N}} \subset  U(1)'   \times \Zf$, both $U(1)$'s are necessary, although $N=4$, $p=2$  is an exception as we will see in the warmup example in Sec.~\ref{wup}.
\begin{figure}[h!]
\centering
\includegraphics[width=1.05\linewidth]{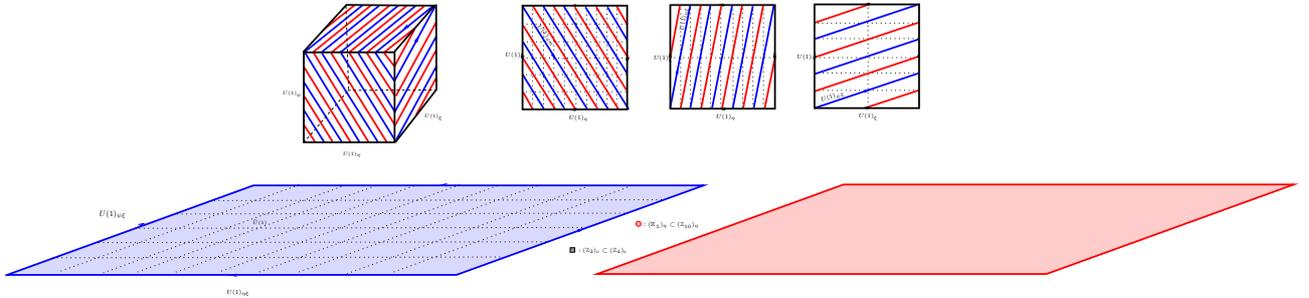}
\caption{\footnotesize The three-torus $U(1)_{\psi} \times U(1)_{\eta} \times U(1)_{\xi}$ broken to $U(1)_{\psi\xi}   \times  U(1)_{\eta\xi}\times (\Z_2)_F$ for the $\{\S,4,2\}$ model. }
\label{S42}
\end{figure}

\end{itemize}

\section{Mixed $({\mathbbm Z}_{2})_F  - [{\mathbbm Z}_{N}]^2$   anomaly      \label{sec:mixedAnom} }


The generalized (mixed) anomaly of the type   $({\mathbbm Z}_{2})_F  - [{\mathbbm Z}_{N}]^2$ was studied in detail in \cite{BKL2} for the $\{\S,N,0\}$   (``$\psi \eta$")  model. 
  We have briefly reviewed the method and results found there  at the end of Introduction. This study is extended below to a wider class of  models
  discussed in Sec.~\ref{models} and Sec.~\ref{symmetries}.   
The global structure of the anomaly-free symmetry group  revealed in Sec.~\ref{symmetries} teaches us that the most interesting class of models 
for the present purpose  are $\{\S,N,p\}$ and  $\{\A,N,p\}$ models with $N$ and $p$ both even, on which our analysis below will set focus.

\subsection{A warmup example $\{\S,4,2\}$  
\label{wup} }

We first consider a simplest, nontrivial model $\{\S,4,2\}$ and  set up the calculation of the mixed anomalies, making a brief note on  
some general features of the gauging of the discrete 1-form  ${\mathbbm Z}_N$  symmetry, on the idea of  ``$({\mathbbm Z}_2)_F$ gauge field",  and 
paying special attention to the way the fermions transform nontrivially under the 1-form ${\mathbbm Z}_N$ gauge transformation.  
The same procedure can then be easily extended
 to more general cases discussed later.

Even though the fact that
\be          {\mathbbm Z}_N   \subset   U(1)_{\psi \xi}  \times U(1)_{\eta \xi} \times   ({\mathbbm Z}_2)_F  \;    \label{formally}  
\ee
has been  proven in general  in Sec.~\ref{ZNinH},     we need an explicit solution for this model, to fix the charges 
of the fermion fields  under the 1-form  $ {\mathbbm Z}_N$  symmetry. 
From 
\be  U(1)_{\psi \xi}: \qquad  \psi: \, e^{i \beta} \;;   \quad \xi:  \, e^{-i \tfrac{N+2}{2} \beta}= e^{-  3  i \beta}    \;;  \nn
\ee
\be  U(1)_{\eta \xi}: \qquad  \eta: \, e^{i \gamma} \;;   \quad \xi:  \, e^{-i \tfrac{N+6}{2} \gamma} =   e^{-  5  i  \gamma}    \;;  \nn
\ee
\be   {\mathbbm Z}_N :   \qquad    \psi: \, e^{4\pi i /N}  =  e^{\pi i } \;;   \quad \eta:  \, e^{-2\pi i /N}  =  e^{- i \pi /2 } \;;   \quad \xi:  \, e^{2\pi i /N}  =  e^{i \pi /2 }\;; \nn
\ee
\be   {\mathbbm Z}_2 :   \qquad    \psi: \, e^{\pm  i \pi }  \;;   \quad \eta:  \, e^{\pm  i \pi } \;;   \quad \xi:  \, e^{\pm  i \pi } \;,   \label{wesee}
\ee
we see that  a simple solution  in this case  is to take  $\beta=0$, and 
$\gamma= + \frac{\pi}{2}\;.$
It is easily seen that  $ {\mathbbm Z}_N$  is realized  as   a  $U_{\eta \xi}(1) \times   ({\mathbbm Z}_2)_F$  transformation with 
\be   {\mathbbm Z}_2 :   \qquad    \psi: \, e^{+ i \pi }  \;;   \quad \eta:  \, e^{- i \pi } \;;   \quad \xi:  \, e^{+ 3 i \pi } \;.\label{chargesz2}
\ee
We  introduce accordingly,  
\begin{itemize}
\item $A$: $ U(1)_{\eta\xi}$    1-form gauge field, 
\item $A_2 $:  $(\mathbb{Z}_{2})_F$ 1-form gauge field, 
\item 
${\tilde a}$: $U(N)_{\rm c} $ 1-form gauge field,
\item $B^{(2)}_\rmc$: $\mathbb{Z}_{N}$ 2-form gauge field.
\end{itemize}
The original $SU(N)$ gauge field $a$  is embedded in a $U(N)$  gauge field  ${\tilde a}$  as  
\be
\widetilde{a}=a+{1\over N}B^{(1)}_\rmc\,, \qquad    N  B_\rmc^{(2)} = d  B_\rmc^{(1)}\;.\label{byone}
\ee
As explained in \cite{GKSW}, \cite{GKKS},  one defines  this way  a globally well-defined $SU(N)/{\mathbbm Z}_N$ connection.  The imposition of the local, 1-form gauge invariance (\ref{continuous1}) below,  eliminates the apparent increase of the degrees
of freedom  (in going from $SU(N)$ to $U(N)$) on the one hand, and at the same time   allows to ``gauge away"  the center ${\mathbbm Z}_N$  variation of Polyakov or Wilson loops 
\be        e^{ i \oint    a }    \to      e^{2\pi i /N} \,  e^{ i \oint    a }   \;, 
\ee   
 on the other.

The 1-form gauge transformation acts on these fields as:
\begin{align}
B^{(2)}_\rmc  &\to B^{(2)}_\rmc+{\diff} \lambda_\rmc\;,  \qquad  \ \,  B^{(1)}_\rmc  \to B^{(1)}_\rmc+{N}\lambda_\rmc\;,  
\nn \\
  {\tilde a} &\to   {\tilde a} +  \lambda_\rmc\;, \qquad   \qquad  {F}({\tilde a})  \to    {F}({\tilde a}) + d \lambda_\rmc;, \label{continuous1}  \\
 A &\to  A -      \lambda_\rmc\;,
  \nn  \\
A_2 &\to A_2+ \frac{N}{2}  \lambda_\rmc =  A_2+ 2 \lambda_\rmc \;.    \label{continuousF} 
\end{align}
As we are here dealing with a  ${\mathbbm Z}_N$  which is a color-flavor locked symmetry  the fermion fields  also  transform as well, 
appropriately.  Their charges 
above follow from Eq.~(\ref{wesee}), Eq.~(\ref{chargesz2}).

 It is perhaps not useless, before proceeding, to remind ourselves of the meaning of a ``$({\mathbbm Z}_2)_F$ gauge field",  $A_2$, which formally looks like an ordinary  $U(1)$ gauge field. Restoring momentarily  the suffices for the  differential forms, 
 \be       2   A_2^{(1)} -   B_\rmc^{(1)}
   =    d   A_2^{(0)}\;  \label{both}
\ee
can be regarded as an  invariant form of the   $({\mathbbm Z}_2)_F$ gauge field,  
 $2   A_2^{(1)} = d   A_2^{(0)}$,   
 where  $A_2^{(0)}$  is a $2\pi$ periodic scalar function (angle). It is an example of an ``almost flat connection":  it satisfies  $2 \, \diff   A_2^{(1)}-{N}B_\rmc^{(2)}=0$
 locally.  However it cannot be set to zero everywhere, as a non vanishing flux through a closed two-dimensional surface may be present, allowing  a nontrivial  $({\mathbbm Z}_2)_F$  holonomy   $  \oint   A_2^{(1)}  =  { 2\pi m}/{2} \;,   \,\,\,     m  \in {\mathbbm Z}$, along a noncontractible closed loop.    A kind of partial gauge fixing would allow us to work with the gauge field $B_\rmc^{(1)}$ and gauge function $\lambda_\rmc$, satisfying always
 \be       \oint  B_\rmc^{(1)} =   2\pi n\;, \qquad    \oint \lambda_\rmc  =  \frac{2\pi \ell}{N}\;,   \qquad \quad  (n  \in {\mathbbm Z}, \quad \ell  \in {\mathbbm Z})\;.
 \ee
 See \cite{BKL2} for more discussions.

The fermion kinetic terms are:  
\bea
&&    \overline{\psi}\gamma^{\mu}\big(\partial +\calR_{\rmS}(\widetilde{a}) -  A_2    \big)_{\mu}P_\rmL\psi + \nonumber\\
&&  \overline{\eta}\gamma^{\mu}\big(\partial + \calR_{\rmF^*}(\widetilde{a})   +  A  +   A_2    \big)_{\mu}P_\rmL\eta +\nonumber \\
&&   \overline{\xi}\gamma^{\mu}\big(\partial + \calR_{\rmF}(\widetilde{a})  - 5 A  -      3  A_2   \big)_{\mu}P_\rmL\xi\;,
\label{naive0}
\eea 
each of which is indeed invariant under  (\ref{continuous1}) and (\ref{continuousF}).
Note that the choice of the  ${\mathbbm Z}_{2}$   charges, ($1, -1, +3$)  for  ($\psi, \eta, \xi$)    fields  (see Eq.~(\ref{chargesz2}))  is dictated by the requirement that 
the redundancy ({\ref{formally}) involving the {\it discrete} symmetries ${\mathbbm Z}_{2}$ and ${\mathbbm Z}_{N}$ be  formally   expressed as an
invariance under (\ref{continuous1})  with a {\it continuous} gauge function  $\lambda_c =  \lambda_c^{\mu}(x) \, dx_{\mu}$.
The 1-form gauge invariant  field tensors are, for the UV fermions $\psi$, $\eta$, $\xi$,  
\bea && \calR_\rmS  F(\tilde{a})  -  {\diff} A_2   \;,      
\nn   \\ && \calR_\rmF^*   F(\tilde{a})  +    {\diff} A      +      {\diff} A_2  \;,
\nn   \\ && \calR_\rmF   F(\tilde{a})   -   5 \,  {\diff} A  -   3   \,   {\diff} A_2     \;.\label{tensors000}
\eea
By rearranging things so that  each term in the bracket is manifestly invariant under  (\ref{continuous1}) and (\ref{continuousF}), this can be rewritten as  
\bea && \calR_\rmS\big(F(\tilde{a})-B^{(2)}_\rmc\big)  -   \left({\diff} A_2 -   2  B^{(2)}_\rmc  \right)\;,       \label{tensors1}
\nn   \\ && \calR_\rmF^*\big(F(\tilde{a})-B^{(2)}_\rmc\big)  +   ( {\diff} A +  B^{(2)}_\rmc)    +      \left({\diff} A_2 -  2  B^{(2)}_\rmc  \right)\;,
\nn   \\ && \calR_\rmF\big(F(\tilde{a})-B^{(2)}_\rmc\big)   -   5  \, ( {\diff} A +  B^{(2)}_\rmc)  -   3 \,  \left( {\diff} A_2 -  2  B^{(2)}_\rmc  \right)  \;.\label{tensors3}
\eea

In the confining vacuum with the full global symmetry, discussed in Sec.~\ref{unbroken},  the infrared degrees of freedom would be  the 
(massless, by assumption)  composite fermions   ${\cal B}_{1}$, ${\cal B}_{2}$, ${\cal B}_{3}$,   (\ref{baryons10}).  Their kinetic terms 
 are   given by  
\bea
&& \overline{{\cal B}_{1}}\gamma^{\mu}\big(\partial +   2 A  +   A_2    \big)_{\mu}P_\rmL {\cal B}_{1}     + \nonumber\\
&&  \overline{{\cal B}_{2}}\gamma^{\mu}\big(\partial -  6  A  -   3 A_2    \big)_{\mu}P_\rmL {\cal B}_{2}  +   \nonumber \\
&& \overline{{\cal B}_{3}}\gamma^{\mu}\big(\partial   + 10   A  +   5  A_2   \big)_{\mu}P_\rmL  {\cal B}_{3}\;.
\label{naive0}
\eea 
The corresponding invariant tensors are 
\bea   && 2  \,( {\diff} A +  B^{(2)}_\rmc)     +      \left[{\diff} A_2  -  2  B^{(2)}_\rmc \right]  \;,  
\nn   \\ && -     6 \, ( {\diff} A +  B^{(2)}_\rmc)       - 3       \left[{\diff} A_2 -  2 B^{(2)}_\rmc \right] \;, 
\nn   \\  &&  10  \, ( {\diff} A +  B^{(2)}_\rmc)     +    5   \left[{\diff} A_2 -  2  B^{(2)}_\rmc \right]  \;, \label{above3}
\eea
respectively.   Though this  formula   appears to depend on   $B^{(2)}_\rmc$ due to the way things have been arranged  to make each term  manifestly  invariant,  $B^{(2)}_\rmc$  actually drops out completely,   reflecting the fact that   ${\cal B}_{1}$, ${\cal B}_{2}$, ${\cal B}_{3}$ are all color $SU(N)$ singlets: there are no gauge kinetic terms in their action.   As a result, there would be  no mixed anomalies in the IR  due to the gauging of ${\mathbbm Z}_N$ 1-form symmetry.

Note that the same cannot be said of the formula  Eq.~(\ref{tensors3})   in the UV theory.   Because,  for instance,  
\be      \tr\,  F(\tilde{a}) =   N\, B^{(2)}_\rmc\;,
\ee
for the fundamental representation,
the $B^{(2)}_\rmc$  dependence of the expressions  in Eq.~(\ref{tensors3})  is not exhausted by the explicit  $B^{(2)}_\rmc$ factors. 
Even though we shall use the formula  Eq.~(\ref{tensors3}) for the calculation of the mixed anomalies below,  for manifest 1-form gauge invariance of our calculation step by step, 
the same  final result can be obtained  (as it should)  by working with a not-term-by-term-manifestly-invariant expression Eq.~(\ref{tensors00}).  This is shown in Appendix~\ref{ofcourse}.
 As a bonus,  the discussion there explains some interesting aspect of our results below.

The rest of the calculations follows that done in \cite{BKL2}.   From Eq.~(\ref{tensors1})    one finds  the $6D$ anomaly functional  in the UV theory \footnote{Even though 
we follow here  the Stora-Zumino descent procedure for  calculating the anomalies,  there is no problem obtaining the same results  \`a la Fujikawa \cite{Fujikawa},  staying in $4D$:  the idea of gauging the center ${\mathbbm Z}_N$ symmetry in itself  has nothing to do with the introduction of the two extra dimensions.   This was explicitly shown  in \cite{BKL2} for the  $\psi\eta$ model. 
}, 
\bea  && {1\over 24\pi^2}\, {\tr}_{\calR_\rmS}\left[\left\{\big( {F}({\tilde a})-B^{(2)}_\rmc\big)  -   \left({\diff} A_2 -  2  B^{(2)}_\rmc  \right)
   \right \}^3   \right]  +     \nonumber \\
   &&  {1\over 24\pi^2}\, {\tr}_{\calR_\rmF^*}\left[\left\{ \big( {F}({\tilde a})-B^{(2)}_\rmc\big)    +   ( {\diff} A +  B^{(2)}_\rmc)    +      \left({\diff} A_2 -  2  B^{(2)}_\rmc  \right)
   \right \}^3   \right]    +     \nonumber \\
 && {1\over 24\pi^2}\, {\tr}_{\calR_\rmF}\left[\left\{\big( {F}({\tilde a})-B^{(2)}_\rmc\big) -   5   ( {\diff} A +  B^{(2)}_\rmc)  -   3   \left( {\diff} A_2 - 2 B^{(2)}_\rmc  \right) 
   \right \}^3   \right]    \;.\label{tensors} 
   \eea  
   Keeping only the relevant terms,  the first line  ($\psi$) gives   
   \be     {1\over 24 \pi^2}  \left[   - 3  (N+2)  {\tr}  \big(F(\tilde{a})-B^{(2)}_\rmc\big)^2     \Big({\diff} A_2 -  2   B^{(2)}_\rmc  \Big)  -   \frac{N(N+1)}{2}   \Big({\diff} A_2 - 2   B^{(2)}_\rmc  \Big)^3
   \right]\;,
   \ee
  the second line  ($\eta$) gives   
  \be     {1\over 24 \pi^2}  \left[  3  (N+6)   {\tr}  \big(F(\tilde{a})-B^{(2)}_\rmc\big)^2     \Big({\diff} A_2 -   \frac{N}{2}   B^{(2)}_\rmc  \Big)  +  N(N+6)   \Big({\diff} A_2 
 -  B^{(2)}_\rmc +\dots  \Big)^3
   \right]\;,
   \ee
   the third line   ($\xi$)  gives:
 \be     {1\over 24 \pi^2}  \left[  -    3\cdot 2 \cdot 3 \,  {\tr}  \big(F(\tilde{a})-B^{(2)}_\rmc\big)^2     \Big({\diff} A_2 -   \frac{N}{2}   B^{(2)}_\rmc  \Big)  +  2N     \Big(  -  3\,  {\diff} A_2 +  B^{(2)}_\rmc  +\ldots \Big)^3
   \right]\;.
   \ee
   Collecting  the relevant terms,  one finds  that  the coefficient of
   \be       {1\over 8 \pi^2}   (B^{(2)}_\rmc)^2\,   {\diff} A_2    
   \ee
   is equal to
   \bea    && N(N+2)-   \frac{N(N+1)}{2}   4   +  (N+6) (-N) + N(N+6)  
     + 2 N   3   +  2N (-3)    \nonumber \\
   &&  = -  N^2 = - 16  \; .  \label{S42model0}
   \eea 
Following the usual procedure   (e.g.,   Eq.~(\ref{usual1}),  Eq.~(\ref{usual2}))  we find  the    mixed  $({\mathbbm Z}_{2})_F  - [{\mathbbm Z}_{N}]^2$    anomaly in $4D$:
\be        -  N^2  \,      {1\over 8 \pi^2}  \int_{\Sigma^4} (B^{(2)}_\rmc)^2    \,   \frac{1}{2}  \delta A_2^{(0)}  = - N^2 \times \frac{\mathbbm Z}{N^2} \,  ({\pm \pi})
=\pm    \pi   \times  {\mathbbm Z}\;.    \label{SNpmodel}
\ee
Namely, the partition function suffers from a sign change under the  fermion parity transformation.
On the other hand,  one would find no $({\mathbbm Z}_{2})_F$ anomaly in the  IR,    if one would assume 
 the chirally symmetric vacuum with the massless baryons    ${\cal B}_{1}$, ${\cal B}_{2}$, ${\cal B}_{3}$ of  Sec.~\ref{unbroken}.
 The contradiction can be avoided by assuming that the system actually  is in a dynamical Higgs phase  such as the one discussed in Sec.~\ref{brokenS}.

\subsection{General $\{\S,N,p\}$ models  with  generic   
$N$ and $p$ even    \label{sec:mixedS}}

Let us now discuss  $\{\S,N,p\}$  systems with general $N$, $p$, both even.
As in the warmup example,   we verify  anew   
\be      {\mathbbm Z}_N \subset  U(1)_{\psi\eta} \times  U(1)_{\psi\xi}  \times {\mathbbm Z}_2\;\label{Z2si}
\ee
 for $N$, $p$ both even, by solving  the equations   \footnote{The charges here are  taken half of those in (\ref{upexBis}).  They would really have be chosen as in Eq.~(\ref{reallycorrect})  
 in order to ensure that the angles $\alpha$ and $\beta$ take the canonical range of $2\pi$,  but the following derivation of the mixed anomaly is not affected by the different choices of the  normalization of the charges and the angles.  }:
\bea 
  \frac{N+ 4 +p}{2}   {\alpha}+   \frac{p}{2} \, {  \beta} &=&  \frac{4\pi}{N} \pm \pi\;,\nonumber \\
  -   \frac{N+2}{2}   {\alpha} &=&   -   \frac{2\pi}{N} \pm \pi\;,\nonumber \\
   -   \frac{N+2}{2}  {\beta}     &=&   \frac{2\pi}{N} \pm \pi\;,  \label{tosolve}
\eea
concretely.   
Indeed,  it is sufficient to find one good solution.   A possible solution is \footnote{This time we first solved the second and third of  Eq.~(\ref{tosolve}), inserted the solutions to the first,  checking that it is indeed satisfied, with appropriate signs for the $({\mathbbm Z}_2)_F$ terms.}
\be  
  {  \alpha}=  \frac{4 \pi}{N(N+2)} + \frac{ 2 \pi}{N+2}\;,   \qquad    {  \beta}=    -   \frac{4 \pi}{N(N+2)}   - \frac{2  \pi}{N+2}\;,\label{possiblesol}
\ee
which is a solution with the  $({\mathbbm Z}_2)_F$ signs    $+ ,  - , + $ for the  $\psi, \eta, \xi$ fields in  Eq.~(\ref{tosolve}), respectively.   
The above solution Eq.~(\ref{possiblesol}) can be simply rewritten as  
\be {  \alpha}=  \frac{ 2\pi}{N}  \;,   \quad    {  \beta}=    - \frac{ 2\pi}{N}   \;.
\label{thatis} 
\ee
As in any anomaly calculation we couple the system to the appropriate background gauge fields, 
\begin{itemize}
\item $A_{\psi\eta}$: $ U(1)_{\psi\eta}$    1-form gauge field, 
\item $A_{\psi\xi}$: $ U(1)_{\psi\xi}$    1-form gauge field, 
\item $A_2 $:  $(\mathbb{Z}_{2})_F
$ 1-form gauge  field, 
\item  $   {\tilde a}$: $U(N)_{\rm c} $ 1-form gauge field, 
\item $B^{(2)}_\rmc$: $\mathbb{Z}_{N}$ 2-form gauge field.
\end{itemize} 
Under the 1-form gauge transformation the fields transform as
\begin{align}
 B^{(2)}_\rmc &\to B^{(2)}_\rmc+{\diff} \lambda_\rmc\;,\qquad \ \, B^{(1)}_\rmc \to B^{(1)}_\rmc+{N}\lambda_\rmc\;, \nn \\ 
     {\tilde a}& \to   {\tilde a} +  \lambda_c\;,  \qquad \qquad \   {\tilde F}({\tilde a}) \to    {\tilde F}({\tilde a}) + d \lambda_c\;, \label{continuous11} \nn \\
  A_{\psi\eta} &\to  A_{\psi\eta}  -      \lambda_c\;,\nn \\  
   A_{\psi\xi}& \to  A_{\psi\xi}  +    \lambda_c\;,\nn \\  
    A_2& \to  A_2+ \frac{N}{2}  \lambda_c \;,
  \end{align}
  where the charges   follow from  (\ref{tosolve}) and  (\ref{thatis}). 
The fermion kinetic terms are:   
\bea
&&   \overline{\psi}\gamma^{\mu}\Big(\partial +\calR_{\rmS}(\widetilde{a})  + \frac{N+4+p}{2}  A_{\psi\eta}  +  \frac{p}{2}  A_{\psi\xi}   +    A_2    \Big)_{\mu}P_\rmL\psi  + \nonumber\\
&& \overline{\eta}\gamma^{\mu}\Big(\partial + \calR_{\rmF^*}(\widetilde{a})  -   \frac{N+2}{2}   A_{\psi\eta}     -   A_2    \Big)_{\mu}P_\rmL\eta+ \nn  \\
 && \overline{\xi}\gamma^{\mu}\Big(\partial + \calR_{\rmF}(\widetilde{a})  -    \frac{N+2}{2}  A_{\psi\xi}  +  A_2 \Big)_{\mu}P_\rmL\xi\;.
\label{naive1}
\eea 
It can be checked readily that each line  is invariant under Eq.~(\ref{continuous11}).  In particular,  the $({\mathbbm Z}_2)_F$ charges are fixed by this requirement. 

The 1-form gauge invariant  field tensors are, for the UV fermions $\psi$, $\eta$, $\xi$,    
\bea &&\!\!\!\!{\cal T}_1 =   \calR_\rmS\big(F(\tilde{a})-B^{(2)}_\rmc\big)    + \frac{N+4+p}{2}  ({\diff}A_{\psi\eta} +  B^{(2)}_\rmc) +  \frac{p}{2}  ({\diff}A_{\psi\xi} -  B^{(2)}_\rmc)      +    \Big({\diff} A_2 -   \frac{N}{2}  B^{(2)}_\rmc  \Big), \label{tensors11111}   \nn \\
   && \!\!\!\!{\cal T}_2 =   \calR_\rmF^*\big(F(\tilde{a})-B^{(2)}_\rmc\big)    -   \frac{N+2}{2}   ({\diff} A_{\psi\eta} +  B^{(2)}_\rmc)  -     \Big({\diff} A_2 -  \frac{N}{2}  B^{(2)}_\rmc  \Big) \;,
\nn \\  && \!\!\! \!{\cal T}_3=  \calR_\rmF\big(F(\tilde{a})-B^{(2)}_\rmc\big)    -   \frac{N+2}{2} ({\diff}A_{\psi\xi} -  B^{(2)}_\rmc)   +     \Big( {\diff} A_2 -   \frac{N}{2}    B^{(2)}_\rmc  \Big) \;,\label{tensors333}
\eea
where appropriate factors of  $B^{(2)}_\rmc$  are  added and subtracted so that each term in the bracket is invariant under  the 1-form gauge transformations (\ref{continuous11}).   
Of course,  the final result does not depend on such a rewriting:   see  Appendix~\ref{ofcourse}.

The $6D$ anomaly functional is   
\be {1\over 24\pi^2}\, \int  \,{\tr}_{\calR_\rmS}   ({\cal T}_1)^3  +  {1\over 24\pi^2}\, \int  \,{\tr}_{\calR_\rmF^*}   ({\cal T}_2)^3  + {1\over 24\pi^2} \, \int  \, {\tr}_{\calR_\rmF}   ({\cal T}_3)^3 \;. 
\ee
Let us now extract the terms relevant to  the  $({\mathbbm Z}_{2})_F  - [{\mathbbm Z}_{N}]^2$ anomaly.
From the $\psi$ contribution one has 
  \be     {1\over 24 \pi^2}  \left[   3  (N+2)  {\tr}  \big(F(\tilde{a})-B^{(2)}_\rmc\big)^2     \Big({\diff} A_2 -   \frac{N}{2}   B^{(2)}_\rmc  \Big)  +  \frac{N(N+1)}{2}   \Big({\diff} A_2 +   2  B^{(2)}_\rmc  +\ldots  \Big)^3
   \right]\;,
   \ee
   $\eta$ gives 
   \be     {1\over 24 \pi^2}  \left[  3  (N+4 +p)   {\tr}  \big(F(\tilde{a})-B^{(2)}_\rmc\big)^2     \Big({\diff} A_2 -   \frac{N}{2}   B^{(2)}_\rmc  \Big)  -  N(N+4 + p)   \Big({\diff} A_2 +  B^{(2)}_\rmc   +\ldots     \Big)^3
   \right]
   \ee
  and  the third line   ($\xi$)  gives:
 \be     {1\over 24 \pi^2}  \left[  3  p \,  {\tr}  \big(F(\tilde{a})-B^{(2)}_\rmc\big)^2     \Big({\diff} A_2 -   \frac{N}{2}   B^{(2)}_\rmc  \Big)  +    p  N     \Big( \,  {\diff} A_2 +  B^{(2)}_\rmc + \ldots   \Big)^3  
   \right]\;.
   \ee
Collecting terms,    one finds  that  the coefficient of
   \be       {1\over 8 \pi^2}   (B^{(2)}_\rmc)^2\,   {\diff} A_2    
   \ee
   is equal to
   \bea   &&  -   N(N+2)  +  \frac{N(N+1)}{2}\cdot 4    - N   (N+4 + p)  + N(N+4 +p)  -   N     p  +    p  N   \nonumber \\     && =    N^2\;.   \label{S42model1}
       \eea 
  A somewhat curious feature of this result (and of   Eq.~(\ref{S42model0})) is that only fermions in a higher representation contribute to the anomaly.  
  The reason for this will become clear in an alternative derivation discussed in Appendix~\ref{ofcourse}.

Following the usual procedure one calculates  the $4D$   mixed  $({\mathbbm Z}_{2})_F  - [{\mathbbm Z}_{N}]^2$  anomaly.    One finds an extra phase in the partition function
associated with the fermion parity transformation in the presence of the ${\mathbbm Z}_{N}$ gauge fields, 
\be          N^2  \,      {1\over 8 \pi^2}  \int_{\Sigma^4} (B^{(2)}_\rmc)^2    \,   \frac{1}{2}  \delta A_2^{(0)}  =    N^2 \times \frac{\mathbbm Z}{N^2} \,  ({\pm \pi})
=\pm    \pi   \times  {\mathbbm Z}\;:
\ee
   there is  a   $({\mathbbm Z}_{2})_F  - [{\mathbbm Z}_{N}]^2$ mixed  anomaly in the theory.

On the other hand,  one finds no $({\mathbbm Z}_{2})_F$ anomaly in the  IR, if one assumes the symmetric vacuum of Sec.~\ref{unbroken}.  This can be seen, as in the warmup example of the previous section,  by simply noting that all  infrared degrees of freedom are color-singlet.   
We conclude that  the chirally symmetric vacuum described by the baryons ${\cal B}_{1}$, ${\cal B}_{2}$, ${\cal B}_{3}$  cannot be realized dynamically.  

We note again that such an inconsistency is avoided, assuming that the system is in the dynamical Higgs phase:   the  color-flavor locked 1-form symmetry  
is spontaneously broken.

\subsection{$\{\A,N,p\}$ models   with  $N$ and $p$ even \label{sec:mixedA}}

The simplest of this class of models,    $\{\A,N,0\}$,  with matter fermions
\be       \yng(1,1) \oplus   (N-4) \,{\bar   {{\yng(1)}}}\;,     \ee
 (``$\chi\eta$  model"),  has been studied, and  the result of the analysis  (unpublished) turns out to be similar to that in the $\psi\eta$ model of \cite{BKL2}, reviewed in Introduction. 
  For even  $N$  the  (nonanomalous) symmetry of the system  
 contains a nonanomalous  $(\mathbb{Z}_2)_F$ factor orthogonal to other continuous symmetry group.  It gets anomalous  under the 1-form gauging of a   ${\mathbbm Z}_N$ center symmetry. 
 This anomaly cannot be reproduced in the infrared, if the vacuum is assumed to be confining, and to keep the full global symmetries.  
 Such a vacuum cannot be realized dynamically.  

Below we study a more general class of   $\{\A,N,p\}$  models, with $p$ additional  pairs of fermions in    $\yng(1) \oplus  {\bar {\yng(1)}}$.  
We check first  
\be      {\mathbbm Z}_N \subset  U(1)_{\chi\eta} \times  U(1)_{\chi\xi}   \times {\mathbbm Z}_2\;.\label{conditionBis}
\ee
Call  $\alpha$  and  $\beta$  the angles associated with  $ U(1)_{\chi\eta}$ and  $U(1)_{\chi\xi}\,$, 
  \bea
&& U(1)_{\chi\eta} : \qquad   \chi\to \rme^{\im \tfrac{N-4+p}{2} \alpha}\chi\;, \quad  \eta \to \rme^{-\im \tfrac{N-2}{2}\alpha}\eta\;, \nn \\
&& U(1)_{\chi\xi} : \qquad   \chi\to \rme^{\im \tfrac{p}{2}  \beta}\chi\;,   \quad  \xi \to \rme^{-\im \tfrac{N-2}{2} \beta}\xi\;.
\label{ucxBis}
\eea  
The condition (\ref{conditionBis})  means that 
\bea   \frac{N- 4 +p}{2}   { \alpha}+ \frac{p}{2} \, {  \beta} &=& \frac{4\pi}{N} \pm \pi\;,\nonumber \\
   - \frac{N-2}{2}  {  \alpha}  &=& -   \frac{2\pi}{N} \pm \pi\;,\nonumber \\
    - \frac{N-2}{2}     {  \beta}    &=&   \frac{2\pi}{N} \pm \pi\;.  \label{tosolveBis}
\eea
It turns out that any two of these imply the third:  there is an arbitrariness to choose from multiple of solutions.    A possible solution is 
\be     {  \alpha}=  \frac{4 \pi}{N(N-2)} -  \frac{2 \pi}{N-2}\;,   \qquad    {  \beta}=    -   \frac{4 \pi}{N(N-2)}  +    \frac{2  \pi}{N-2}\;,\label{possiblesolisBis}
\ee
which is a solution with the  $({\mathbbm Z}_2)_F$ signs in  Eq.~(\ref{tosolveBis}),    $-\pi,  +\pi, - \pi$ for the  $\chi, \eta, \xi$ fields, respectively.   
Actually the solution Eq.~(\ref{possiblesolisBis}) is,  simply,   
\be  {  \alpha}= - \frac{2 \pi}{N}    \;,   \qquad    {  \beta}=     \frac{ 2\pi}{N} \;. \label{thatisBis} 
\ee
The color-flavor locked    $ {\mathbbm Z}_N $  transformation,
(\ref{tosolveBis}) and  (\ref{thatisBis}), together with the normalization of  the 1-form gauge field $\lambda_c$,  
fix the charges of the fermion fields  in Eq.~(\ref{naive2})  below.  

We introduce the background gauge fields
\begin{itemize}
\item $A_{\chi\eta}$: $ U(1)_{\chi\eta}$    1-form gauge field, 
\item $A_{\chi\xi}$: $ U(1)_{\chi\xi}$    1-form gauge field, 
\item $A_2 $:  $(\mathbb{Z}_{2})_{F}$ 1-form gauge field
 field, 
\item   $  {\tilde a}$: $U(N)_{\rm c} $ 1-form gauge field, 
\item $B^{(2)}_\rmc$: $\mathbb{Z}_{N}$ 2-form gauge field.
\end{itemize}
Under the 1-form gauge transformation
\begin{align}
B^{(2)}_\rmc &\to B^{(2)}_\rmc+{\diff} \lambda_\rmc\;,  \qquad  \ \, B^{(1)}_\rmc \to B^{(1)}_\rmc+{N}\lambda_\rmc\;,  \nn \\
{\tilde a} &\to   {\tilde a} +  \lambda_c\; , \qquad \qquad \  {\tilde F}({\tilde a}) \to    {\tilde F}({\tilde a}) + d \lambda_c\;, \label{continuous100} 
\nn   \\ 
 A_{\chi\eta} &\to  A_{\chi\eta}  +    \lambda_c\;,\qquad   \nn   \\  
 A_{\chi\xi} &\to  A_{\chi \xi}  -      \lambda_c\;,\qquad  \nn   \\  
A_2 &\to A_2+ \frac{N}{2}  \lambda_c \;.
\end{align}
The fermion kinetic terms are:   (the charges   follow from    (\ref{thatisBis}))
\bea
&&   \overline{\chi}\gamma^{\mu}\Big(\partial +\calR_{\rmA}(\widetilde{a})  +   \frac{N-4+p}{2}  A_{\chi\eta}  +  \frac{p}{2}  A_{\chi\xi}    -    A_2    \Big)_{\mu}P_\rmL\chi +  \nonumber\\
&& \overline{\eta}\gamma^{\mu}\Big(\partial + \calR_{\rmF^*}(\widetilde{a})  -   \frac{N-2}{2}   A_{\chi\eta}     +   A_2    \Big)_{\mu}P_\rmL\eta +  \nn  \\
&&    \overline{\xi}\gamma^{\mu}\Big(\partial + \calR_{\rmF}(\widetilde{a})   -      \frac{N-2}{2}  A_{\chi\xi}  -     A_2 \Big)_{\mu}P_\rmL\xi\;. 
\label{naive2}
\eea 
It is seen that each line  is invariant under (\ref{continuous100}).  In particular,  the $({\mathbbm Z}_2)_F$ charges are fixed by this requirement. 

The 1-form gauge invariant  field tensors are, for the UV fermions $\chi$, $\eta$, $\xi$,    
\bea
 &&  \!\!\!\!{\cal T}_1 =  \calR_\rmA\big(F(\tilde{a})-B^{(2)}_\rmc\big)   +  \frac{N- 4+p}{2}  ({\diff}A_{\chi\eta} -  B^{(2)}_\rmc) +  \frac{p}{2}  ({\diff}A_{\chi\xi} +  B^{(2)}_\rmc)     -    \Big({\diff} A_2 -   \frac{N}{2}  B^{(2)}_\rmc  \Big) ,      \label{tensors11111}\nn \\
 &&  \!\!\!\! {\cal T}_2  = \calR_{\rmF^*}\big(F(\tilde{a})-B^{(2)}_\rmc\big)    -   \frac{N-2}{2}   ({\diff} A_{\chi\eta} -   B^{(2)}_\rmc)  +   \Big({\diff} A_2 -  \frac{N}{2}  B^{(2)}_\rmc  \Big)\;,\nn \\
  &&  \!\!\!\! {\cal T}_3  =  \calR_\rmF\big(F(\tilde{a})-B^{(2)}_\rmc\big)    -     \frac{N-2}{2} ({\diff}A_{\chi\xi} +  B^{(2)}_\rmc)   -    \Big( {\diff} A_2 -   \frac{N}{2}    B^{(2)}_\rmc  \Big)  \;. \label{tensors33333}
\eea
The $6D$ anomaly functional is   
\be {1\over 24\pi^2}\, \int  \,{\tr}_{\calR_\rmA}   ({\cal T}_1)^3  +  {1\over 24\pi^2}\, \int  \,{\tr}_{\calR_{\rmF^*}}   ({\cal T}_2)^3  + {1\over 24\pi^2} \, \int  \, {\tr}_{\calR_\rmF}   ({\cal T}_3)^3 \;. 
\ee
Let us now extract the terms relevant to  the  $({\mathbbm Z}_{2})_F  - [{\mathbbm Z}_{N}]^2$ anomaly.
From the $\chi$ contribution one has 
  \be     {1\over 24 \pi^2}  \left[  - 3  (N-2)  {\tr}  \big(F(\tilde{a})-B^{(2)}_\rmc\big)^2     \Big({\diff} A_2 -   \frac{N}{2}   B^{(2)}_\rmc  \Big)  -     \frac{N(N-1)}{2}   \Big({\diff} A_2 -   2  B^{(2)}_\rmc  +\ldots  \Big)^3
   \right]\;,
   \ee
   $\eta$ gives 
   \be     {1\over 24 \pi^2}  \left[  3  (N- 4 +p)   {\tr}  \big(F(\tilde{a})-B^{(2)}_\rmc\big)^2     \Big({\diff} A_2 -   \frac{N}{2}   B^{(2)}_\rmc  \Big)  +  N(N-4 + p)   \Big({\diff} A_2 -  B^{(2)}_\rmc   +\ldots     \Big)^3
   \right]
   \ee
   and  the third line   ($\xi$)  gives:
 \be     {1\over 24 \pi^2}  \left[  -    3  p \,  {\tr}  \big(F(\tilde{a})-B^{(2)}_\rmc\big)^2     \Big({\diff} A_2 -   \frac{N}{2}   B^{(2)}_\rmc  \Big) -  p  N     \Big( \,  {\diff} A_2 - B^{(2)}_\rmc + \ldots   \Big)^3  
   \right]\;.
   \ee
Collecting terms,    one finds  that  the coefficient of
   \be       {1\over 8 \pi^2}   (B^{(2)}_\rmc)^2\,   {\diff} A_2    
   \ee
   is equal to
   \bea   &&    N(N-2)  -  \frac{N(N-1)}{2}\cdot 4    +  (N- 4 + p) (-N) + N(N-4 +p)   +  N   p  -    p  N   \nonumber \\    \label{ANpmodel}
    &&     =    -  N^2\;.
       \eea 
Following the usual procedure one calculates  the $4D$   mixed  $({\mathbbm Z}_{2})_F  - [{\mathbbm Z}_{N}]^2$  anomaly,
\be       -  N^2  \,      {1\over 8 \pi^2}  \int_{\Sigma^4} (B^{(2)}_\rmc)^2    \,   \frac{1}{2}  \delta A_2^{(0)}  =  N^2 \times \frac{\mathbbm Z}{N^2} \,  ({\pm \pi})
=\pm    \pi   \times  {\mathbbm Z}\;.
\ee
That is, the partition function changes sign under  the fermion parity, $\chi, \eta, \xi \to -\chi, -\eta, -\xi $.     
 In other words,  we found  a   $({\mathbbm Z}_{2})_F  - [{\mathbbm Z}_{N}]^2$  mixed  anomaly in the UV theory.

On the other hand,  one finds no $({\mathbbm Z}_{2})_F$ anomaly in the  IR, assuming 
 the chirally symmetric vacuum with the massless baryons ${\cal B}_{1}$, ${\cal B}_{2}$, ${\cal B}_{3}$.   This then cannot be the correct phase of the system.

\section{Summary  \label{conclude}}

In this work we have extended the study of mixed anomalies affecting a chiral discrete $({\mathbbm Z}_2)_F$ symmetry,  found  \cite{BKL2} 
in a simple chiral gauge theory ($\psi\eta$ model),   to a wider class of models, the general Bars-Yankielowicz and the generalized Georgi-Glashow models.  

   Writing the effects of instantons on the  three $U(1)$'s associated with the three fermions as
     \be
U(1)_{\psi} \times U(1)_{\eta}   \times U(1)_{\xi}  \xrightarrow{\rm anomaly}   {\cal H} \;,
\ee
the global symmetry of these models  $G_{\mathrm{f}}$  can be written,  for $\{\S,N,p\}$ models, for instance,    as 
\be   G_{\mathrm{f}} \xrightarrow{\rm anomaly}\frac{  SU(N+4+p) \times  SU(p)  \times   {\cal H} }{ \mathbb{Z}_{N}  \times \mathbb{Z}_{N+4+p} \times \mathbb{Z}_{p}} \; 
\ee
and similarly for $\{\A,N,p\}$ models,  with a replacement,  $N+4+p \to  N-4+p$.   The division by various centers has been explained in Sec.~\ref{symmetries}. 

  In both classes of the models,  if one of  $N$ and $p$  (or both)  is  odd,   ${\cal H}$, hence $G_f$, has a connected structure.    It can be taken as 
 \be    {\cal H} =  U(1)_{1}   \times  U(1)_{2}   \;,  
 \ee
 where  $U(1)_{1,2}$ are arbitrary two of the  nonanomalous combinations,  $U(1)_{\psi\eta}$, $U(1)_{\psi\xi}$, and  $U(1)_{\xi\eta}$. 
It follows that,   once   the conventional anomaly matching equations are all satisfied with respect to  $G_F$,  
considering the mixed anomalies involving the 1-form   discrete center symmetry  $\mathbb{Z}_{N}$  does not provide us with any new information about the 
candidate phase of the system.     
The UV-IR matching involving any new, mixed anomalies  is a simple consequence of (i.e., included in)  the conventional anomaly matching equations. 
This is similar to what was found in \cite{BKL2} for odd $N$ $\psi\eta$ models.

For this reason, the main part of our analysis here has been focused on the models with $N$ and $p$, both even.   In all cases of this type,   the global symmetry  $G_f$ has two, disconnected components, as
\be    {\cal H} =  U(1)_{1}   \times  U(1)_{2}  \times  ({\mathbbm Z}_2)_F  \;.  
 \ee
 $({\mathbbm Z}_2)_F$  is nonanomalous, as all other factors in  $G_F$,  but the fact that it is nonanomalous hinges upon  the integer instanton numbers 
\be    \frac{1}{8\pi^2}  \,  \left(  \int_{\Sigma_4}   \tr  \,F^2   \right)  \in   {\mathbbm Z}    \ee
and  is not a simple result of an algebraic cancellation of the contributions from 
different fermions, as is the case for the continuous, nonanomalous symmetries $   U(1)_{\psi\xi}   \times  U(1)_{\eta\xi}$.   
This can be checked by inspecting   Eqs.~(\ref{naive0}), ~(\ref{naive1}) and  (\ref{naive2}).    For instance,    in the warmup example of  the $\{\S,4,2\}$ model,  the effect of
the chiral transformations,    
\be     \psi \to   e^{-i \pi} \psi\;,   \quad   \eta \to   e^{i  \pi} \eta\;, \quad    \xi \to   e^{- 3  i \pi} \xi\;, 
\ee
(see Eq.~(\ref{naive0}))   is the extra  phase in the partition function
\be      \{ -(N+2) + (N+6) - 3 \cdot 2 \}  \,   \frac{1}{8\pi^2}  \,  \left(  \int_{\Sigma_4}  \tr \,  F^2   \right)    \cdot      \pi    =  -2  \pi  \,  {\mathbbm Z} \;:   \label{indeedirr} 
\ee
which is indeed irrelevant, but only because the instanton numbers are quantized to integers.   
 The nonanomalous  $({\mathbbm Z}_2)_F$   symmetry  
has  thus a different status as compared to other, continuous nonanomalous symmetries such as   $ U(1)_{\chi\eta},\,  U(1)_{\chi\xi}\,$ and $U(1)_{\eta\xi} $.

But this means that, once   all fields are coupled to  the 1-form  center ${\mathbbm Z}_N$ gauge fields   $\big(B_\rmc^{(2)},  B_\rmc^{(1)}\big)$
\be  N  B_\rmc^{(2)} = d  B_\rmc^{(1)}\;,   
\ee
and fractional 't Hooft  fluxes are allowed,    a mixed    $({\mathbbm Z}_{2})_F  - [{\mathbbm Z}_{N}]^2 $ anomaly may arise. 
  In other words,   there may be an obstruction against gauging  the 1-form  center ${\mathbbm Z}_N$ symmetry and $0$-form  
$({\mathbbm Z}_2)_F$ symmetry  simultaneously.    

Our calculations show  that  such an obstruction   (a generalized 't Hooft anomaly) is indeed  present.

On the other hand,  such an obstruction could not occur in the chirally symmetric confining vacuum of  Sec.~\ref{unbroken}, as the infrared fermions are all singlets of $SU(N)$.   Consistency requires that either the assumption of confinement or  that of  unbroken global symmetry  (no condensates),  or both,   must be abandoned.  

There is no inconsistency in the other, possible vacua in the infrared (dynamical Higgs phase,  Sec.~\ref{brokenS} and Sec.~\ref{brokenA}),   as   
$ U(1)_{\chi\eta},\,  U(1)_{\chi\xi}\,$ and $U(1)_{\eta\xi} $  are broken spontaneously by the condensate, so is the color-flavor locked  1-form center ${\mathbbm Z}_N$ symmetry.  

Note that the 0-form  $({\mathbbm Z}_2)_F$  symmetry itself does not need to be, and indeed is not,  spontaneously broken, since all bifermion condensates are invariant under   
\be   \psi, \eta, \xi \to    -\psi, -\eta,  -\xi \;. 
 \ee   
In fact, as this  fermion parity
 coincides with an angle $2\pi$ space rotation,  a spontaneous breaking of   $({\mathbbm Z}_2)_F$  would have meant the spontaneous breaking of the Lorentz invariance, which does not occur.
 
   In this respect,  even though  the mixed anomaly   $({\mathbbm Z}_{2})_F  - [{\mathbbm Z}_{N}]^2$  found in \cite{BKL2} and confirmed  here for  an extended class of models, looks similar at first sight  to the mixed anomaly 
   $CP -    [{\mathbbm Z}_{N}]^2$    found recently  \cite{GKKS}  in the pure $SU(N)$ Yang-Mills theory at $\theta = \pi$,   the {\it way} the mixed anomaly manifests
itself in the infrared physics is different.  In the latter case,   the new anomaly is consistent with,  or implies,   the phenomenon of the double vacuum degeneracy   and the consequent spontaneous $CP$ breaking  \cite{Dashen},  which was known from the QCD Effective Lagrangian analysis \cite{DiVecchiaVenez,Witten:1980sp}  and also from soft supersymmetry breaking perturbation \cite{Konishi,Evans} of  the exact Seiberg-Witten solutions \cite{SW1,SW2} of pure ${\cal N}=2$  supersymmetric Yang-Mills theory.
 
In our case, the  mixed anomaly   $({\mathbbm Z}_{2})_F  - [{\mathbbm Z}_{N}]^2$ means instead  that confinement and the full global chiral symmetries (no condensates)  are incompatible: one or both must be abandoned.   The dynamical Higgs phase discussed in Sec.~\ref{brokenS},  Sec.~\ref{brokenA},  seems to be fully consistent with this requirement.

Before concluding,  let us add a few more clarifying remarks. The first concerns the interpretation of our analysis. In spite of the presence of fermions in the fundamental (or antifundamental) representation, our system classically has an exact color-flavor locked 1-form  ${\mathbbm Z}_{N}$ symmetry, as in (\ref{ZNpsieta}), (\ref{ZNequiv}) for the $\psi\eta$ model and similar equivalence relations for other models.  We have decided to gauge this 1-form symmetry, by introducing appropriate two-form gauge field  $B_\rmc^{(2)}$ and gauge fields  $A$ (for $U_{\psi\eta}(1)$)  and $A_2$ (for ${\mathbbm Z}_2$),  by requiring  the invariance under 1-form gauge transformations (\ref{werequire}) and (\ref{simult}).  Similarly for all other models.  This fixed the form of all the fermion and gauge kinetic terms. We find that, actually, there is an anomaly of mixed terms  $({\mathbbm Z}_{2})_F  - [{\mathbbm Z}_{N}]^2$,  and it means that there is an obstruction to such a gauging. By definition this is a 't Hooft anomaly,
of a new kind.  The rest is as in the standard 't Hooft anomaly analysis. We require the same anomaly be present in the infrared. We find that  if the system is in confining, flavor symmetric vacua  with no condensates and with a set of gauge invariant, massless composite fermions as the only infrared degrees of freedom, this anomaly ``matching" fails, as the low-energy theory cannot have the same anomaly.  On the other hand,  in a dynamical Higgs phase with bifermion condensates,  the $U(1)$ symmetry which is part of the  color-flavor locked ${\mathbbm Z}_{N}$ symmetry, is broken spontaneously.  One may argue that 
this is a spontaneous breaking of the ${\mathbbm Z}_{N}$ center symmetry,  but the usual association of  
unbroken (broken) center symmetry with confinement (Higgs) phase in pure Yang-Mills theory, is perhaps not quite adequate here.  

The second remark concerns an alternative possibility for the infrared system. When a discrete symmetry is broken spontaneously by the gauge dynamics,  it is sometimes possible that the infrared system is described by a TQFT.  However, in the models studies here, there are massless degrees of freedom (NG bosons and/or massless fermions) which are required to be present to reproduce the anomalous and nonanomalous comtinuous symmetries of the underlying systems.  Thus our systems cannot be a pure TQFT in the infrared, in that simple sense. 
However, we have not excluded the possibility that a TQFT plays a more subtle role in the infrared physics of our modes, e.g., coupled to the massless baryons and somehow reproduces the mixed  $({\mathbbm Z}_{2})_F  - [{\mathbbm Z}_{N}]^2$ anomaly.   This question will be left for a future investigation.

 To conclude, the analysis presented here confirms that the result found in \cite{BKL2} - that an extended symmetry consideration
implies a dynamical Higgs phenomenon in a class of chiral gauge theories -
 is not an accidental one peculiar to  the simplest  models considered there,
but holds true in a much larger class of theories. Such a result should, in our view,  be regarded as
  a rather  {\it general} property  of strongly-coupled chiral gauge theories.

\section*{Acknowledgments}

This work is supported by the INFN special research project grant  ``GAST"  (Gauge and String Theories).

\appendix

\section{The mixed   $({\mathbbm Z}_{2})_F  - [{\mathbbm Z}_{N}]^2 $ anomaly: a master formula  \label{ofcourse}}

In this Appendix, we show that our results on the mixed anomaly  $({\mathbbm Z}_{2})_F  - [{\mathbbm Z}_{N}]^2 $  found in  Sec.~\ref{sec:mixedAnom} do not depend on the rearrangement of the fermion tensors to term-by-term manifestly invariant form,  as done  in Eq.~(\ref{tensors3}),  Eq.~(\ref{above3}),
Eq.~(\ref{tensors333}), and  Eq.~(\ref{tensors33333}).   The result of this discussion is a sort of master formula, which better express certain aspects of our analysis.

For concreteness, let us first  take the warmup example of  Sec.~\ref{wup}.
The  $6D$   anomaly functional is, from (\ref{tensors00}),    
\bea  && {1\over 24\pi^2}\, {\tr}_{\calR_\rmS}\left[\left\{F(\tilde{a})  -  {\diff}  A_2
\right \}^3   \right]  +     \nonumber \\
&&  {1\over 24\pi^2}\, {\tr}_{\calR_\rmF^*}\left[\left\{F(\tilde{a})    +    {\diff} A    +     {\diff} A_2
\right \}^3   \right]    +     \nonumber \\
&& {1\over 24\pi^2}\, {\tr}_{\calR_\rmF}\left[\left\{F(\tilde{a}) -   5\, {\diff} A  -   3 \, {\diff} A_2 
\right \}^3   \right]    \;.\label{cosa} 
\eea  
For the purpose of finding  the $({\mathbbm Z}_2)_F$ anomaly,  
we expand these, and integrate once to  find  the   $5D$  WZW action proportional to $A_2$.   The variation of the form  
 \be    \delta A_2  =   \frac{1}{2}   \de \,  \delta A_2^{(0)}\;, \qquad    \delta A_2^{(0)} = \pm 2\pi\;,   
   \ee 
 then  leads to an anomalous surface term -   the anomaly in $4D$  theory -   given by the phase 
 \be    \frac{1}{8\pi^2}\int_{\Sigma_4}   {\cal  P}  \, \, \frac{\delta A_2^{(0)}}{2} \;, \qquad   \frac{\delta A_2^{(0)}}{2}=\pm \pi
 \ee
 where
\bea  {\cal P}    &=&       -{\tr}_{\calR_\rmS}\left[F(\tilde{a})^2\right]     +(N+p+4)\,  {\tr}_{\calR_\rmF^*}\left[F(\tilde{a})^2\right]
\nn \\
&-&     3 \, p\,  {\tr}_{\calR_\rmF}\left[F(\tilde{a})^2\right]   \;,
\eea
($N=4$, $p=2$), where
  the trace taken in a representation ${R}$  is indicated  by ${\tr}_{R}$.    Now
\bea  
{\tr}_{R}   \left[\big(F(\tilde{a})\big)^2\right] &=&  {\tr}_{R}\left[\big(F(\tilde{a}) - B^{(2)}_c + B^{(2)}_c\big)^2\right]    \nonumber\\
&=& {\tr} \left[ \mathcal{R}_R     \big( F(\tilde{a}) - B^{(2)}_c  \big)     +   {\cal N}(R) B^{(2)}_c    {\mathbbm 1}_{d(R)}       \right]^2  \nonumber\\
&=&{\tr}   \left[ \mathcal{R}_R     \big(F(\tilde{a}) - B^{(2)}_c)^2   + {\cal N}(R)^2    \big(B^{(2)}_c \big)^2   {\mathbbm 1}_{d(R)}     \right]  \;,
\eea
where   $\mathcal{R}_R$ indicates the  matrix form appropriate for the representation $R$,  ${\cal N}(R)$ its $N$-ality,    and use was made of the fact that 
\be     {\tr}_{R}   \big(F(\tilde{a}) - B^{(2)}_c \big)  = 0\;,
\ee
valid for an $SU(N)$  element in any representation.  $ {\mathbbm 1}_{d(R)} $ stands for  the  $d(R)\times d(R)$   unit matrix,  where $d(R)$ is the dimension of the representation $R$.  
 Calculating the above, one finds
\bea
 {\tr}_{R}   \left[\big(F(\tilde{a})\big)^2\right] &=&   D(R)\,  {\tr}_{F}\left[\big(F(\tilde{a}) - B^{(2)}_c\big)^2\right] + d(R)  {\cal N}(R)^2\big(B^{(2)}_c\big)^2 =\nonumber\\
&=& D(R)\,   {\tr}_{F}\left[F({\tilde a}) \right]^2 + \left[  - D(R) \cdot N +   d(R)   {\cal N}(R)^2  \right]  \big(B^{(2)}_c\big)^2\;,  \label{result}
\eea
where  $D(R)$ is twice  the Dynkin index  $T_R$,  
\be   {\rm {tr}} \big( t_R^a t_R^b \big) =  T_R  \,  \delta^{ab}\,,
\ee
normalized as 
\be       T_R= \frac{1}{2}\;, \qquad  D(R)  =1\;, \qquad    R= \yng(1)  \,\,  {\rm or} \,\,  {\bar {\yng(1)}} 
\;.
\ee

Now  
\be      \frac{1}{8\pi^2}\int_{\Sigma_4}     {\tr}_{F}\left[F({\tilde a})^2 \right]    \in {\mathbbm Z},  
\ee
and the first term in Eq.~(\ref{result})  corresponds to the conventional instanton contribution to the  $({\mathbbm Z}_2)_F$ anomaly,  which is known to be absent
(for instance, see Eq.~(\ref{indeedirr})) \footnote{The combination 
\[ \frac{1}{8\pi^2}   \int_{\Sigma_4}    \{   {\tr}   {\tilde F}^2 -  {\tr}   {\tilde F} \wedge  {\tr}   {\tilde F} \}  \]
is the second Chern number of $U(N)$ and is an integer.  The second term of the above is also an integer. 
}.


Therefore   the nonvanishing  mixed  $({\mathbbm Z}_2)_F-  {\mathbbm Z}_N^2$  anomaly comes only from  the second term of  Eq.~(\ref{result}),
containing the 2-form gauge field.   One finds  that the total   $({\mathbbm Z}_2)_F-  {\mathbbm Z}_N^2$  anomaly  is given by
\be \boxed{ \Delta S^{({\rm Mixed}\, {\rm anomaly})}  = (\pm \pi) \cdot  \sum_{fermions}   c_2  \,  (d(R)  {\cal N}(R)^2- N \, D(R))     \,     \frac{1}{8\pi^2}\int_{\Sigma_4}       \big(B^{(2)}_c\big)^2 \;.   }      \label{generalres}
\ee
$c_2$  are the   ${\mathbbm Z}_2$ charges  (the coefficients of  ${\diff} A_2$  in (\ref{cosa}) in the example of Sec.~\ref{wup}).  
This is our master formula.

%
Applying this  formula to the $\{{\cal S},4,2\}$ model of Sec.~\ref{wup}, Eq.~(\ref{cosa}),  one gets   ($\pm \pi$ times)  
\bea  &&
\frac{1}{8\pi^2} \int_{\Sigma_4}    \left\{   -    \left(4 \cdot \frac{N(N+1)}{2} - N(N+2)\right) + 10\left(N -N\right) -6\left(N -N\right) \right\} \big(B^{(2)}_c\big)^2  \nonumber\\
&=&   \frac{-N^2}{8\pi^2} \int  \big(B^{(2)}_c\big)^2  \;,     
\eea
which is indeed  the result found in Sec.~\ref{wup}.

 Note that for $R=F$  (the fundamental) or  $R=F^*$  (antifundamental),    $d(R)=N$, ${\cal N}(R)=D(R)=1$,  therefore 
 \be  d(R) {\cal N}(R)^2- N \cdot D(R)) =0\;, \qquad   ( R= \yng(1)  \,\,  {\rm or} \,\,  {\bar {\yng(1)}} )\;:
 \ee
 these fermions do not 
 contribute to the   $({\mathbbm Z}_{2})_F  - [{\mathbbm Z}_{N}]^2$ mixed anomaly.
 And this explains a somewhat curious feature in the results observed  earlier in  Eq.~(\ref{S42model0}), Eq.~(\ref{SNpmodel}) and Eq.~(\ref{ANpmodel}).

 The  formula (\ref{generalres}) is valid for a fermion in a generic representation, so it  can be applied  at once to  the general   $\{\S,N,p\}$  and  $\{\A,N,p\}$ models, yielding  an extra phase in the partition function
 under the fermion parity, 
\be
\Delta S=\frac{\pm \pi}{N^2} \left(d(\S)\cdot {\cal N}(\S)^2-N\cdot D(\S)\right)= \frac{\pm\pi}{N^2} \left(\frac{N(N+1)}{2}\cdot 4   - N(N+2)\right)=\pm\pi \;,
\ee
 for the $\{\S, N, p\}$ model,   and 
\be
\Delta S=\frac{\pm \pi}{N^2} \left(d(\A)\cdot {\cal N}(\A)^2-N\cdot D(\A)\right)= \frac{\pm \pi}{N^2} \left(\frac{N(N-1)}{2} \cdot 4 - N(N-2)\right)=\pm\pi\;,
\ee
for the $\{\A, N, p\}$ model,   in agreement with the results found  in  Sec.~\ref{sec:mixedS} and in  Sec.~\ref{sec:mixedA}.

\end{document}